\documentclass{article}

\usepackage{arxiv}
\usepackage{subcaption}
\usepackage[utf8]{inputenc} 
\usepackage[T1]{fontenc}    
\usepackage{hyperref}       
\usepackage{url}            
\usepackage{booktabs}       
\usepackage{amsfonts}       
\usepackage{nicefrac}       
\usepackage{microtype}      
\usepackage{lipsum}
\usepackage{graphicx}
\usepackage{graphicx}
\usepackage{dcolumn}
\usepackage{bm}
\usepackage[utf8]{inputenc}
\usepackage[T1]{fontenc}
\usepackage{mathptmx}
\usepackage{etoolbox}
\usepackage{algorithm}
\usepackage{algorithmic}
\usepackage{hyperref}
\usepackage{xcolor}
\usepackage{ulem}
\usepackage[utf8]{inputenc}
\usepackage[T1]{fontenc}
\usepackage{mathptmx}
\usepackage{etoolbox}
\usepackage{amsmath}
\usepackage{amsfonts}
\usepackage{amssymb}
\definecolor{asparagus}{rgb}{0.33, 0.42, 0.18}

\graphicspath{ {./images/} }

\title{Girsanov Reweighting for Uncertainty Propagation in Rare-Event Kinetics}

\usepackage{authblk}

\author[1,2]{Léonard Moracchini}
\author[1,*]{Thomas Pigeon}
\author[1]{Morgane Menz}
\author[1]{Thibault Faney}
\author[3,4]{Thomas D. Swinburne}
\author[2]{Mihai-Cosmin Marinica}

\affil[1]{IFP Energies nouvelles (IFPEN), 1--4 Av. du Bois Préau, 92852 Rueil-Malmaison, France}
\affil[2]{Université Paris-Saclay, CEA, SRMP, 91191 Gif-sur-Yvette, France}
\affil[3]{Department of Mechanical Engineering, University of Michigan, Ann Arbor, MI 48109, USA}
\affil[4]{Aix-Marseille Université, CNRS, CINaM UMR 7325, 13288 Marseille, France}
\affil[*]{Corresponding author: \texttt{thomas.pigeon@ifpen.fr}}
\begin{document}
\maketitle


\begin{abstract}
Machine-learning interatomic potentials (MLIPs) have become a powerful tool for rare event sampling in molecular dynamics, offering near \textit{ab initio} accuracy at a fraction of the computational cost. However, the uncertainty associated with these models remains a major challenge. Existing uncertainty quantification approaches have largely focused on point-wise quantities, such as energies and forces, or on equilibrium thermodynamic observables. In this work, we introduce a framework for propagating MLIP uncertainty to the averaged committor probability, a kinetic observable that enables reaction-rate calculations. Our approach combines rare event sampling methods such as Adaptive Multilevel Splitting with Girsanov reweighting to estimate the sensitivity of committor probabilities to variations in MLIP parameters, without requiring the costly resampling of reactive trajectories for each parameter realization. We derive full and approximate Girsanov-based estimators for uncertainty propagation and validate them on several benchmark systems, including a rugged Müller–Brown potential, a dimer in a solvent, and the conformational transition of butane. The proposed framework enables the construction of uncertainty-aware probability distributions for rare event observables and successfully recovers reference rare event probabilities from uncertain surrogate models. Under mild assumptions on the accuracy of the MLIP within metastable basins, the framework can also provide uncertainty bounds on reaction rates through Hill's relation. These results demonstrate that path-space reweighting provides an efficient route for propagating MLIP uncertainty to rare event kinetics.

\end{abstract}

\section{Introduction}

The exact dynamics of systems of interest in chemistry and biology are often difficult, if not impossible, to fully characterize with a purely experimental approach. Indeed, some intermediate states are too short-lived for instruments to capture any signature of their presence. Consequently, modeling these systems at the atomic scale based on fundamental physics laws has become a crucial complementary tool to describe \textit{in silico} the features missing from experimental observations. In this context, a standard framework is Molecular Dynamics (MD), where the electronic degrees of freedom are integrated out to define an interatomic potential energy surface $V(\mathbf{x})$ via first-principles calculation at fixed atomic positions $\mathbf{x}$. The time evolution of the nuclei is then governed by classical equations of motion determined by $V$. In cases relevant for chemical or biological applications, the potential $V$ presents a complex landscape with multiple basins corresponding to the various metastable states of the system. The system remains trapped within a given state $A$ for a long time before eventually undergoing a fluctuation large enough to cross a high energetic or entropic barrier to reach another metastable state $B$. Mathematically, the transition timescale $\tau_{AB}$ is orders of magnitude larger than the characteristic vibrational period of the system~\cite{cossio_transition_2018}, meaning that the probability of observing a reactive trajectory (a path leaving $A$ and entering $B$ without returning to $A$) within standard MD simulations vanishes. This highlights a fundamental paradox: while the transition event itself is short-lived, the waiting time to observe it is prohibitively long. Since the fastest atomic oscillations dictate a discretization time-step $\Delta t \sim 1 \;\mathrm{fs}$, computing the transition rate constant $k_{AB}$ via brute-force MD becomes computationally intractable.

Various methodologies have been developed to circumvent this timescale separation; Transition State Theory (TST)~\cite{vanden-eijnden_transition_2005, hanggi1990fifty} remains the most widely employed. Grounded in equilibrium statistical mechanics, TST assumes a clear separation of timescales and evaluates the flux across a predefined dividing surface. Yet, it often lacks reliability when complex, entropic, or multi-channel paths co-exist, and it fundamentally fails to account for recrossing events or to provide an ensemble of reactive trajectories~\cite{gesvandtnerova_importance_2024}.
Consequently, alternative frameworks rooted in Transition Path Theory (TPT)~\cite{e_towards_2006, e_transition-path_2010} have emerged that focus specifically on reactive path sampling. 
The Hill relation~\cite{hill2012free} transforms the computation of $k_{AB}$ into a rare event probability computation via the committor probability $\langle p_{A\to B}\rangle_{ \lambda_{\partial{A}}}$ averaged against the exit distribution $\lambda_{\partial{A}}$ defined on the boundary $\partial A$ of $A$.
It corresponds to the probability for a trajectory, when starting from the reactant boundary $\partial A$, to hit $B$ before reaching $A$. Various methods were designed to compute $k_{AB}$ and $\langle p_{A\to B}\rangle_{ \lambda_{\partial{A}}}$ by sampling reactive trajectories. While Transition Path Sampling (TPS)~\cite{dellago_transition_1998} and Transition Interface Sampling (TIS)~\cite{van2003novel} sample trajectories in path space through local Monte Carlo perturbations, particle splitting techniques such as Forward Flux Sampling~\cite{allen_FFS_2009} (FFS), Sequential Monte Carlo~\cite{del_moral_feynman-kac_2004} (SMC), and Adaptive Multilevel Splitting~\cite{cerou_AMS_2007} (AMS) leverage an interacting particle system where poorly performing replicas are iteratively discarded and replaced by branches of better-adapted trajectories. The latter type of methods, on which we will focus, has been successfully used for the sampling of rare events in a wide range of applications, including catalysis~\cite{pigeon_unbiased_2025}, protein folding~\cite{borrero_folding_2006}, and protein-ligand dissociation~\cite{teo2016adaptive}.

To alleviate the computational cost inherent to these long simulations, MLIPs trained to reproduce a given level of first-principles calculations results, most often density functional theory (DFT), have been increasingly used in molecular dynamics. Although these potentials achieve impressive force and energy errors, they still require robust uncertainty quantification (UQ). Significant research has addressed this challenge in recent years, largely focusing on epistemic uncertainty to identify training data scarcity within configurational space, thereby facilitating active learning~\cite{kulichenko_uncertainty-driven_2023,carrete_deep_2023}. More recently, attention has shifted toward model misspecification errors~\cite{swinburne_parameter_2025,perez_uncertainty_2025}, that characterize the under-parameterization of the model. While reference data (DFT energies and forces) exhibit negligible aleatoric noise, machine learning models remain susceptible to inherent structural limitations. For instance, MLIPs can exhibit dynamical instabilities despite low point-wise errors~\cite{fu_forces_2023}, alongside severe transferability failures outside their training manifold~\cite{montes_de_oca_zapiain_training_2022}. Furthermore, model distillation inevitably compounds misspecification errors, as evidenced by the error curves plotted against training set size and parameter count across diverse models~\cite{mazitov_pet-mad_2025}. Many popular UQ methodologies~\cite{grasselli_uncertainty_2025} used in MD focus on the values (i.e forces and energies), such as conformal prediction~\cite{ho2026flexible}, mixture of experts~\cite{moe_1,moe_2} and ensemble methods~\cite{carrete_deep_2023, kellner_uncertainty_2024}, but other frameworks also provide a proper posterior distribution over the parameter space, like Bayesian regression~\cite{hegde_bayesian_2024}, Laplace approximation~\cite{bigi_prediction_2024} or Point-wise Optimal Parameter Set (POPS) method~\cite{swinburne_parameter_2025, perez_uncertainty_2025}. The latter type of methods is particularly advantageous, as it enables the propagation of uncertainty through sequential simulations to final derivative observables. However, efficient propagation relies on calculating observables independently of the specific parametrization of the interatomic potential, as recently achieved for vibrational free energy~\cite{swinburne_agnostic_2025}. Yet, traditional UQ methods do not directly solve the problem of propagating these parameter uncertainties onto rare event kinetic observables, such as the averaged committor probability $\langle p_{A\to B}\rangle_{ \lambda_{\partial{A}}}$ or the transition rate $k_{AB}$.

Reweighting methodologies provide a powerful framework for mapping observables computed under a specific reference ensemble (characterized by a given potential energy surface or temperature) onto a target state. These approaches generally fall into two categories~\cite{keller_dynamical_2024}: thermodynamic reweighting, which applies to equilibrium properties, and dynamical reweighting, which applies to kinetic observables. The former relies solely on configurational phase-space averages, making it independent of the actual time-evolution of the system. Conversely, reweighting a dynamical observable is considerably more challenging, as it requires evaluating weight corrections over entire path trajectories. Recently, Girsanov theorem-based reweighting has emerged as a mathematically rigorous and promising approach to handle such path-space trajectories~\cite{donati_girsanov_2017,kieninger_girsanov_2023,kieninger_path_2021}. Reweighting techniques are particularly valuable in the context of enhanced sampling~\cite{keller_dynamical_2024}, where the underlying potential or temperature is biased to increase the frequency of rare events. Furthermore, while these frameworks have also been applied to uncertainty propagation, their scope has been strictly limited to thermodynamic observables so far~\cite{imbalzano_uncertainty_2021}. By bridging this gap, this work leverages Girsanov dynamical reweighting to propagate point-wise uncertainties to the averaged committor probability $\langle p_{A\to B}\rangle_{ \lambda_{\partial{A}}}$.

The remainder of this article is organized as follows. Section~\ref{sec:theory} establishes the theoretical framework, encompassing rare event sampling techniques, the mathematical formalism for uncertainty quantification and propagation in molecular dynamics, and Girsanov reweighting principles. Section~\ref{sec:method} derives the proposed Girsanov-based estimators for rare event probabilities, specifically formalizing the full Girsanov estimator and an approximate but computationally tractable cumulant estimator. Section~\ref{sec:sensitivity} validates these estimators via numerical experiments on a dimer in a Weeks-Chandler-Andersen~\cite{weeks1971role} (WCA) solvent, a rugged extended Müller-Brown potential~\cite{muller1979location}. Section~\ref{sec:results_up} applies the complete methodology to an academic example and to a physical system, quantifying the uncertainty of conformational rare event probabilities in butane using MACE foundation models~\cite{batatia2023foundation} and the POPS UQ scheme. Finally, Section~\ref{sec:conclusion} provides concluding remarks.
\begin{figure}[t!]
    \centering
    \includegraphics[width=\linewidth]{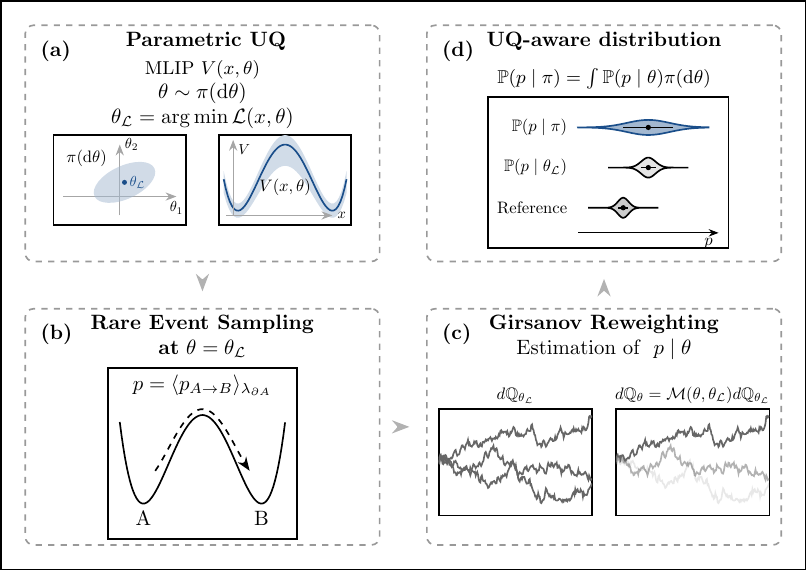}
    \caption{\textbf{Schematic pipeline of parametric uncertainty propagation for rare event probabilities.} 
\textbf{(a)} Uncertainty quantification of the MLIP potential parameters $\theta$ with respect to the posterior distribution $\pi(\mathrm{d}\theta)$ centered at the Maximum Likelihood Estimation (MLE) $\theta_\mathcal{L}$.
\textbf{(b)} Evaluation of the committor probability $p = \langle p_{A\to B}\rangle_{ \lambda_{\partial{A}}}$ at $\theta = \theta_\mathcal{L}$ characterizing the system's transition from the reference state $A$ to the target state $B$.
\textbf{(c)} Path-space reweighting via Girsanov's theorem, mapping the nominal probability measure $\mathrm{d}\mathbb{Q}_{\theta_\mathcal{L}}$ to the perturbed measure $\mathrm{d}\mathbb{Q}_{\theta}$.
\textbf{(d)} Resulting uncertainty-aware probability distribution $\mathbb{P}(p \mid \pi)$ (blue curve) capturing parametric uncertainties, compared against the MLE profile $\mathbb{P}(p|\theta_\mathcal{L})$ and the reference profile $\mathbb{P}_{\mathrm{ref}}(p)$.}
    \label{fig:schema}
\end{figure}
\section{Theoretical Framework}\label{sec:theory}
We consider a system of $N$ particles in a periodic box of volume $\mathcal{V}$ at temperature $\mathcal{T}$. The evolution of the particle positions $\mathbf{x} \in \Omega \subset \mathbb{R}^{3N}$ and momenta $\textbf{p} \in \mathbb{R}^{3N}$ is dictated by the Stochastic Differential Equation (SDE) commonly referred to as underdamped Langevin dynamics: 
\begin{equation}
	\label{eq:underdamped_langevin_dynamics}
	\left\{ 
	\begin{aligned}
		\mathrm{d}\textbf{x}_t & = M^{-1}\textbf{p}_t \, \mathrm{d}t, \\
		\mathrm{d}\textbf{p}_t & = -\nabla V(\textbf{x}_t) \, \mathrm{d}t - \zeta \textbf{p}_t \,\mathrm{d}t + \sqrt{\frac{2 \zeta}{\beta}} M^{\frac{1}{2}}\mathrm{d} \, \textbf{B}_t,
	\end{aligned} \right.
\end{equation}
where the friction coefficient $\zeta > 0$ has the dimension of an inverse time, $M$ is the mass matrix, $V:\Omega\to \mathbb{R}$ is a potential and $\textbf{B}_t$ is a standard $3N$-dimensional Brownian motion. 
In the limit of $\zeta \to \infty$, after an appropriate rescaling of time ($t \to \zeta t$), the inertia term becomes negligible, yielding the overdamped Langevin dynamics:
\begin{equation}
    \label{eq:overdamped_langevin_dynamics}
    \mathrm{d}\mathbf{x}_t = -\nabla V(\mathbf{x}_t) \, \mathrm{d}t + \sqrt{\frac{2}{\beta}} \, \mathrm{d}\mathbf{B}_t.
\end{equation}

For pedagogical clarity, the theoretical framework and the derivation of the Girsanov estimators in Sections~\ref{sec:theory} and~\ref{sec:method} are presented within the simpler framework of overdamped Langevin dynamics. However, since actual molecular dynamics simulations often require accounting for inertial effects, all numerical benchmarks presented in Sections~\ref{sec:sensitivity} and \ref{sec:results_up} are conducted using underdamped Langevin dynamics, using the explicit extension detailed in Appendix~\ref{app:OBABO}.

\subsection{Rare event sampling methods}
A common need in MD is to compute the rare event probability of transitioning between a reactant state $A$ and a product state $B$. Let us consider the committor function, which represents the probability for a trajectory $X = (\mathbf{x}_t)_{t \ge 0}$ starting in $\mathbf{x}$ to reach $B$ before $A$:
\begin{equation}
p_{A\to B}(\mathbf{x}) = \mathbb{P} ^\mathbf{x}(\tau_B < \tau_A),
\end{equation}
where $\tau_A$ and $\tau_B$ are the first hitting times of the sets $A$ and $B$.

The committor function can be used to identify collective variables~\cite{jung2019artificial,chen2023discovering,frohlking2025learning}, but it also serves to compute reaction rates, which is the main focus of this work. Indeed, the Hill relation~\cite{hill2012free, Lelievre2024} relates the reaction rate $k_{AB}$ to the flux $\phi_A$ of trajectories leaving state $A$ and the committor function averaged against the exit distribution $\lambda_{\partial{A}}$  on a boundary surface $\partial A$:
\begin{equation}\label{hill_relation}
    k_{AB} = \phi_A \, \langle p_{A\to B}\rangle_{\lambda_{\partial{A}}}.
\end{equation}
In the overdamped setting, $\partial A$ is defined as a boundary layer located at a small distance $\varepsilon$ from the reactant state to avoid the trivial vanishing of the committor function on $\partial A$. It is a fixed hyperparameter of the Hill relation that is assumed to be constant throughout the rest of the paper. Meanwhile, the exit distribution $\lambda_{\partial{A}}$ is given by the Gibbs measure restricted to $\partial A$:
\begin{equation*}
    \lambda_{\partial{A}}(\mathbf{x}) \propto e ^{-\beta V(\mathbf{x})}\sigma_{\partial A}(\mathrm{d}\mathbf{x}),
\end{equation*}
where \(\sigma_{\partial A}\) denotes the surface measure on \(\partial A\).
Throughout the remainder of this work, this rare event probability will be denoted by $p$:
\begin{equation}
p \equiv \langle p_{A\to B}\rangle_{\lambda_{\partial{A}}} = \int_{\partial A} p_{A \to B}(\mathbf{x}) \,\lambda_{\partial{A}}(d\mathbf{x}).
\end{equation}

Several sampling methods based on multilevel splitting have been developed to compute this probability, such as  Forward Flux Sampling (FFS)~\cite{allen_FFS_2009}, Sequential Monte-Carlo (SMC)~\cite{del_moral_feynman-kac_2004}, and  Adaptive Multilevel Splitting (AMS)~\cite{cerou_AMS_2007}.  These methods provide unbiased estimators $\hat{p}_\mathrm{spl}$ of the rare event probability and a collection of $N_T$ reactive trajectories $\{X_1, \ldots, X_{N_T}\}$ starting in $\partial A$ and ending in $B$. These algorithms exhibit a nonzero variance, denoted by $\sigma_\mathrm{spl}^2$, which depends on the quality of the chosen reaction coordinate as well as on the number of replicas employed. In practice, since the sampling algorithms satisfy Central Limit Theorem (CLT)~\cite{cerou2019adaptive}, $n_{\mathrm{real}}$ independent simulations $\{\hat{p}^{(1)}_\mathrm{spl}, \ldots, \hat{p}^{(n_{\mathrm{real}})}_\mathrm{spl}\}$ are computed, resulting in a law on the empirical mean $\bar{p}_\mathrm{spl}$:
\begin{equation}\label{distribution_ams}
    \bar{p}_\mathrm{spl} \equiv \frac{1}{n_\mathrm{real}} \sum _{i=1}^{n_\mathrm{real}}\hat{p}_\mathrm{spl}^{(i)}\sim \mathcal{N}\left( p, \, \frac{\sigma_\mathrm{spl}^2}{n_{\mathrm{real}}} \right).
\end{equation}
For certain algorithms, the properties of the path distributions have also been established~\cite{del_moral_feynman-kac_2004,charles-edouard_unbiasedness_2015, cerou2019adaptive}. For instance, for SMC and AMS, it has been shown that the empirical distribution $\hat{\gamma}$:
\begin{equation}\label{hat_gamma}
    \hat{\gamma} = \frac{\hat{p}_\mathrm{spl}}{{N_T}} \sum_{i=1}^{N_T}\delta_{X_i},
\end{equation}
where $\delta_{X_i}$ denotes the Dirac measure on the path space concentrated at $X_i$, is an unbiased estimator of the distribution on the path space $\gamma$, defined for any test function $\varphi$ by:
\begin{equation}\label{eq}
\gamma(\varphi) = \mathbb{E}\left[\varphi(X) \mathbf{1}_{{\tau_B < \tau_A}}\right].
\end{equation}
The unbiasedness of the path distribution estimator $\hat{\gamma}$ is central to our work, as it allows, through an appropriate choice of $\varphi$, the reweighting of the paths sampled under the initial potential to a different target potential. 
In the remainder of this paper, we focus on the AMS algorithm ($\mathrm{spl} = \mathrm{ams}$). 
This choice is motivated by its ability to minimize variance through the automatic and adaptive positioning of the intermediate surfaces~\cite{cerou2019adaptive}. 
A comprehensive description of the AMS method, which is used in this study, is provided in Appendix~\ref{app:ams}.
It is worth noting, however, that our framework remains agnostic to the specific choice of the underlying sampling method.
\subsection{Uncertainty Quantification  for Molecular Dynamics}\label{sec:uq_linear}
Uncertainty Quantification for Molecular Dynamics is a rapidly growing field~\cite{grasselli_uncertainty_2025}. The most commonly employed strategies are ensemble methods and their variants. The central concept consists in training the MLIP $V(\mathbf{x},\theta)$, with parameters $\theta \in \mathbb{R}^p$, on the same dataset using different initializations of the parameter vector. This procedure yields a collection of distinct optimal parameter sets $\{\theta_1, \ldots, \theta_{n_e}\}$, where typical ensemble sizes are in the range $n_e \in [5, 20]$~\cite{grasselli_uncertainty_2025,kellner_uncertainty_2024}. Despite their simplicity, recent studies have shown that ensemble methods can fail to capture the true epistemic uncertainty in the case of Graph Neural Networks~\cite{vieira_deep_2026}. Furthermore, they do not yield a realistic posterior distribution over the parameter space, as it is a sum of Dirac distributions:
$$\pi_{\mathrm{ensemble}}(\theta) = \frac{1}{n_e}\sum_{i=1}^{n_e} \delta(\theta -\theta_i) $$
To enable rigorous propagation of uncertainty, it is necessary to specify an appropriate distribution for the parameters. In general, the exact posterior distribution over the parameters of a MLIP based on neural networks is intractable due to the size of the models. A standard solution consists of focusing on a subset of the model parameters. This has been achieved, for instance, by considering only the final layer of the network, an approach known as the Last-Layer Prediction Rigidity~\cite{bigi_prediction_2024}. Recently~\cite{swinburne_agnostic_2025,perez_uncertainty_2025}, it was proposed to exploit a linear parametrization with respect to the per-atom descriptors $\mathbf{D}_i : \Omega \to \mathbb{R}^D$ of the model. This is achieved by formulating the MLIP in a linear form:
\begin{equation}
\label{eq:general_lin_form}
  V^L (\mathbf{x}, \theta) \stackrel{\mathrm{def}}{=} \theta^\top \mathcal{D}(\mathbf{x}), \quad \mathcal{D}(\mathbf{x}) \stackrel{\mathrm{def}}{=}\sum_{i=1} ^N \hat{\phi}(\mathbf{D}_i(\mathbf{x})),  
\end{equation}
where $\hat{\phi}(\mathbf{D}_i)$ is a $P$-dimensional featurization of per-atom descriptors $\mathbf{D}_i$. This framework includes several linear-in-descriptors models, where $\hat{\phi}(\mathbf{D}_i) = \mathbf{D}_i$, like SOAP~\cite{bartok2013representing}, SNAP~\cite{thompson2015spectral}, ACE~\cite{lysogorskiy2021performant}, MILADY~\cite{goryaeva2021efficient},  as well as potentials using  polynomial or kernel featurizations, like qSNAP~\cite{rohskopf2023fitsnap} or GAP~\cite{bartok2015g}. It can also encompass architectures that are not globally linear in their parameters by restricting attention to a subset of trainable parameters. For instance, in the Message-Passing Neural Network \texttt{MACE} architecture~\cite{batatia2023foundation}, taking the input of the final readout layer as $\mathbf{D}_i$, we can write the featurization $\hat{\phi}_{\text{MACE}}$ as:
\begin{equation}
    \hat{\phi}_{\text{MACE}} (\mathbf{D}_i) = \mathbf{D}_i \oplus f(\mathbf{D}_i),
\end{equation}
where $f$ is a frozen one-layer neural network. Quantifying model misspecification by focusing exclusively on this linear subset of parameters may provide a conservative upper bound on the uncertainty. Intuitively, restricting the optimization to a subset of parameters reduces the flexibility of the model and therefore tends to overestimate the structural error relative to the full architecture. This approach has been successfully evaluated on the \texttt{mptraj} dataset~\cite{perez_uncertainty_2025}. The linear parameterization of foundational MLIPs can also be leveraged for transfer learning~\cite{novelli2025fast}.

Using this framework, one can easily apply UQ methods designed for linear models, such as Bayesian ridge regression~\cite{hegde_bayesian_2024,kurniawan2022bayesian}. However, as demonstrated in recent work~\cite{swinburne_parameter_2025}, this framework is not suited for molecular dynamics due to the near-deterministic nature of ab initio data (e.g., Density Functional Theory), which leads to vanishing parameter uncertainty. To address this limitation, the Point-wise Optimal Parameter Set (POPS) framework was recently introduced~\cite{swinburne_parameter_2025}. POPS is robust to the non-aleatoric nature of the data and provides a posterior distribution that explicitly accounts for the structural misspecification error of the model. The POPS scheme has been tested across a wide variety of observables and datasets~\cite{swinburne_parameter_2025, perez_uncertainty_2025,ho2026flexible}, yielding a posterior distribution $\pi_\mathrm{POPS}$ over the parameter space. We give more details about the limitations of classical Bayesian Regression and the derivation of the POPS method in Appendix~\ref{app:POPS}. Although the numerical examples in this paper use the parameter distribution $\pi_\mathrm{POPS}$, the proposed framework is applicable to any parameter distribution $\pi$.

\subsection{Uncertainty Propagation}\label{sec:uncertainty_propagation}
We aim to quantify the distribution of the observable $p = \langle p_{A\to B}\rangle _{\lambda_{\partial{A}}}$ under a parameter distribution $\pi(\mathrm{d}\theta)$, which corresponds to the conditional distribution~\cite{dai_uncertainty_2025} $\mathbb{P}(p \mid \pi)$:
\begin{equation}
    \mathbb{P}(p \mid \pi) = \int_{\mathbb{R}^p} \mathbb{P}(p \mid \theta) \pi(\mathrm{d}\theta),
\end{equation}
with $\mathbb{P}(p\mid \theta)$ the distribution of the conditional law $p\mid \theta$. When parameter uncertainty is neglected, i.e., the distribution is a Dirac measure at the maximum-likelihood estimate (MLE) parameter $\theta_\mathcal{L}$, $\pi_\mathrm{MLE}(\mathrm{d}\theta) = \delta(\theta - \theta_{\mathcal{L}})\mathrm{d}\theta$, the uncertainty reduces strictly to the sampling uncertainty: $\mathbb{P}(p \mid \pi_\mathrm{MLE}) = \mathbb{P}(p \mid \theta_{\mathcal{L}})$, which can be calculated with equation \eqref{distribution_ams}. Conversely, if model parameters are uncertain, $\pi(\mathrm{d}\theta)$ is non-trivial and the conditional distribution $\mathbb{P}(p \mid \theta)$ must be estimated. 

A straightforward approach~\cite{imbalzano_uncertainty_2021, wen_uncertainty_2020}, valid for any generic observable, consists of sampling parameter sets $\{\theta_1, \ldots, \theta_{n_e}\}$ from $\pi(\mathrm{d}\theta)$ and running multiple independent simulations of the algorithm to obtain a standard Monte Carlo estimate:
\begin{equation}
    \mathbb{P}(p \mid \pi) \approx \frac{1}{n_e}\sum_{i=1}^{n_e} \mathbb{P} (p \mid \theta_i).
\end{equation}

\noindent While this strategy is inherently compatible with popular ensemble-based methods, it becomes computationally prohibitive when each evaluation of the algorithm to compute $p\mid \theta_i$ is expensive. To alleviate this burden, efficient approximation methods must be deployed to evaluate this expectation across the parameter space with only one expensive probability computation at the MLE parameter $p \mid \theta_\mathcal{L}$, which is the aim of the next section.

\subsection{Girsanov Reweighting}\label{sec:girsanov_reweighting}

We refer to Girsanov reweighting~\cite{kieninger_girsanov_2023} as the use of path probability ratios to reweight dynamical properties computed under a modified potential. This approach is rooted in the works of Onsager and Machlup~\cite{OnsagerMachlup1953} and Girsanov~\cite{Girsanov1960}. Girsanov reweighting can be interpreted as an importance sampling method over path space. 
Recently, Girsanov reweighting has been coupled with enhanced sampling techniques~\cite{keller_dynamical_2024} and used to identify reaction coordinates~\cite{shmilovich_girsanov_2023}. Girsanov reweighting has also been applied for sensitivity analysis (SA) and uncertainty quantification (UQ) purposes~\cite{pantazis_relative_2013, tsourtis_parametric_2015}, notably by computing the relative entropy rate between two potential parameterizations. This allows for the derivation of bounds on certain thermodynamic quantities under an equilibrium measure. However, to the best of our knowledge, no sensitivity analysis for kinetic quantities, such as the committor probability, has been reported in the literature. An intuitive exposition of Girsanov reweighting is presented in this subsection.

Let us consider a path  $X = (\mathbf{x}_0,\mathbf{x}_1,\ldots,\mathbf{x}_{n_{\tau}})$ sampled up to a stopping time $\tau$,  with a time step $\Delta t$. For overdamped Langevin dynamics, the Euler-Maruyama integration scheme under a reference potential $V$ is given by:
\begin{equation}
    \mathbf{x}_{k+1} = \mathbf{x}_k - \nabla V(\mathbf{x}_k)\Delta t + \sqrt{2\beta^{-1}\Delta t}\,\xi_k, \quad \xi_k \sim \mathcal{N}(0, I).
\end{equation}
The transition probability density $\pi_\mathrm{dyn}(\mathbf{x}_{k+1}|\mathbf{x}_k)$ is Gaussian:
\begin{equation}
    \pi_\mathrm{dyn}(\mathbf{x}_{k+1}|\mathbf{x}_k) \propto \exp\left( - \frac{\beta}{4\Delta t} \left\| \mathbf{x}_{k+1} - \mathbf{x}_k + \nabla V(\mathbf{x}_k)\Delta t \right\|^2 \right).
\end{equation}
Under a target perturbed potential $\tilde{V} = V + U$, the transition density $\tilde{\pi}_\mathrm{dyn}(\mathbf{x}_{k+1}|\mathbf{x}_k)$ follows the same functional form with $V$ replaced by $\tilde{V}$. The reweighting factor for the complete discrete trajectory is obtained via the likelihood ratio:
\begin{align}
    \frac{\mathrm{d}\mathbb{Q}_{\tilde{V}}(X)}{\mathrm{d}\mathbb{Q}_V(X)} &= \frac{\mu_{\tilde{V}}(\mathbf{x_0})}{\mu_{V}(\mathbf{x_0})} \prod_{k=0}^{n_\tau-1} \frac{\tilde{\pi}_\mathrm{dyn}(\mathbf{x}_{k+1}|\mathbf{x}_k)}{\pi_\mathrm{dyn}(\mathbf{x}_{k+1}|\mathbf{x}_k)} \nonumber ,\\
    &= \frac{\mu_{\tilde{V}}(\mathbf{x_0})}{\mu_{V}(\mathbf{x_0})}\exp \Bigl( - \frac{\beta}{4\Delta t} \sum_{k=0}^{n_\tau-1} \Big[ \| \mathbf{x}_{k+1} - \mathbf{x}_k + \nabla \tilde{V}(\mathbf{x}_k)\Delta t \|^2 \nonumber  - \| \mathbf{x}_{k+1} - \mathbf{x}_k + \nabla V(\mathbf{x}_k)\Delta t \|^2 \Big] \Big) \nonumber,
\end{align}
where $\mu_{\tilde{V}}$ and $\mu_V$ refer to the probability measure on the initial point induced by the potentials.
By expanding the squares and substituting $\mathbf{x}_{k+1} - \mathbf{x}_k + \nabla V(\mathbf{x}_k)\Delta t = \sqrt{2\beta^{-1}\Delta t}\,\xi_k$, the terms quadratic in $\nabla V$ cancel, yielding:
\begin{eqnarray}
\frac{\mathrm{d}\mathbb{Q}_{\tilde{V}}(X)}{\mathrm{d}\mathbb{Q}_V(X)} &=& \frac{\mu_{\tilde{V}}(\mathbf{x_0})}{\mu_{V}(\mathbf{x_0})}\exp \Biggl( - \sum_{k=0}^{n_\tau-1} \left[ \sqrt{\frac{\beta}{2}} \nabla U(\mathbf{x}_k) \cdot \xi_k \sqrt{\Delta t} \right. \nonumber \\
& & \left. + \frac{\beta}{4} \|\nabla U(\mathbf{x}_k)\|^2 \Delta t \right] \Biggr).
\label{eq:likelihood_overdamped}
\end{eqnarray}
This expression, which constitutes the discretized version of Girsanov's theorem~\cite{Girsanov1960}, is particularly powerful. Indeed, it allows for the exact evaluation of the likelihood ratio of a given path under two distinct potentials. This computation only requires additionally storing the realization of the driving noise $(\xi_k)$ and evaluating the difference between the forces derived from the two potentials along the sampled trajectory.

A well-known limitation of Girsanov-based path reweighting is the exponential growth of its variance with respect to both the trajectory duration and the magnitude of the potential perturbation~\cite{keller_dynamical_2024, wang_marginal_2025}. This scaling behavior typically renders the approach computationally intractable for long equilibrium simulations or substantial potential discrepancies. 
In our framework, this issue is mitigated by two factors. First, we apply Girsanov reweighting to reactive trajectories sampled via splitting methodologies, whose duration is significantly shorter than the mean first-passage time~\cite{cossio_transition_2018}. Second, the potential energy differences arising from the uncertainty quantification of the machine learning model remain sufficiently bounded across the configurational space explored by the reactive ensemble, thereby preventing pathological variance inflation.

\section{Method}\label{sec:method}

We consider Girsanov reweighting around a parameterized potential $V(\mathbf{x},\theta)$, where we wish to vary the parameter $\theta$ around a reference value $\theta_{\mathcal{L}}$. Both the exit distribution $\lambda_{\partial{A}}^\theta$ on $\partial A$ and the committor function $p_{A\to B}^\theta$ now depend on the parameter $\theta$ ($\partial A$ itself is a hyperparameter of the Hill relation and does not change with $\theta$). Let $p(\theta)$ denote the true committor probability, averaged over the reactant boundary $\partial A$ with distribution $\lambda_{\partial{A}}^\theta$, for the potential $V(\mathbf{x},\theta)$:
\begin{equation}
    p(\theta) = \int_{\partial A} p_{A\to B} ^\theta(\mathbf{x}) \lambda_{\partial{A}}^\theta(\mathrm{d}\mathbf{x})
\end{equation}
Let $\hat{p}_{\mathrm{ams}}(\theta_{\mathcal{L}})$ denote the AMS estimator of the committor probability $p(\theta_{\mathcal{L}})$ at the reference parameter $\theta_{\mathcal{L}}$, and let $\{X_1,\ldots,X_{N_T}\}$ denote the sampled reactive trajectories. The aim of this section is to build estimators of the parametric probability $p(\theta)$  using only trajectories sampled at $\theta_\mathcal{L}$, and to derive an approximation of the law $p\mid \theta$. For this purpose, we use the Girsanov reweighting framework.

\subsection{General case}
Let us consider a path  $X = (\mathbf{x}_0,\mathbf{x}_1,\ldots,\mathbf{x}_{n_{\tau}})$ sampled up to a stopping time $\tau$, with an Euler-Maruyama scheme, Gaussian noise terms $\{\xi_0, \ldots, \xi_{n_\tau-1}\}$ and discretization step $\Delta t$.
The likelihood ratio between two path-space probability measures $\mathbb{Q}_{\theta_{\mathcal{L}}}$ and $\mathbb{Q}_{\theta}$ for the path $X$ can be computed using equation~\ref{eq:likelihood_overdamped}. Let us denote this quantity $\mathcal{M}(X, \theta_\mathcal{L,}\theta)$:

\begin{align}\label{eq:girsanov_weight}
    \mathcal{M}(X, \theta_\mathcal{L},\theta)
    &= \frac{\mathrm{d}\mathbb{Q}_{\theta}(X)}
            {\mathrm{d}\mathbb{Q}_{\theta_{\mathcal{L}}}(X)} \nonumber \\
    &= 
    \frac{\mathrm{d}\lambda_{\partial{A}}^{\theta}}
         {\mathrm{d}\lambda_{\partial{A}}^{\theta_{\mathcal{L}}}}(\mathbf{x}_0)
    \nonumber 
    \exp \Biggl(
    - \sum_{k=0}^{n_\tau-1}
    \left[
    \sqrt{\frac{\beta}{2}}
    \nabla \left[V(\mathbf{x}_k,\theta)-V(\mathbf{x}_k,\theta_\mathcal{L})\right]
    \cdot \xi_k \sqrt{\Delta t}
    \right. \nonumber \\
    &\qquad\qquad\left.
    +
    \frac{\beta}{4}
    \left\|
    \nabla \left[V(\mathbf{x}_k,\theta)-V(\mathbf{x}_k,\theta_\mathcal{L})\right]
    \right\|^2
    \Delta t
    \right]
    \Biggr).
\end{align}
with
${\mathrm{d}\lambda_{\partial{A}}^{\theta}}
     /{\mathrm{d}\lambda_{\partial{A}}^{\theta_{\mathcal{L}}}}$
the Radon--Nikodym derivative of the initial distribution on
$\partial A$ associated with the potential parameter $\theta$
with respect to the reference distribution associated with
$\theta_{\mathcal{L}}$. In the overdamped setting, where the exit distribution is the Gibbs measure restricted to $\partial A$, it is given by:
\begin{align}
  \frac{\mathrm{d}\lambda_{\partial{A}}^{\theta}}
       {\mathrm{d}\lambda_{\partial{A}}^{\theta_{\mathcal{L}}}}(\mathbf{x}_0)
  &=
  \frac{\exp[-\beta(V(\mathbf{x}_0,\theta)-V(\mathbf{x}_0,\theta_{\mathcal{L}}))]}
       {\mathbb{E}_{\lambda_{\partial{A}}^{\theta_{\mathcal{L}}}}
       [\exp[-\beta(V(\mathbf{x},\theta)-V(\mathbf{x},\theta_{\mathcal{L}}))]]}.
\end{align}

Using equation~\ref{eq:girsanov_weight}, one can define the full Girsanov estimator $\hat{p}_\mathrm{G}(\theta)$ of $p(\theta)$ using the $N_T$ trajectories generated under the reference potential $V(\mathbf{x}, \theta_{\mathcal{L}})$ as:
\begin{align}\label{p_g}
    \hat{p}_\mathrm{G}(\theta) &\stackrel{\mathrm{def}}{=} \hat{p}_{ \mathrm{ams}}(\theta_\mathcal{L}) \cdot \frac{1}{{N_T}} \sum_{i=1}^{N_T} \mathcal{M}(X_i, \theta_\mathcal{L},\theta).
\end{align}
By noticing that $\hat{p}_\mathrm{G}(\theta)=\hat{\gamma}(\mathcal{M}(\cdot, \theta_\mathcal{L},\theta))$, where $\hat{\gamma}$ is defined in equation~\ref{hat_gamma}, it is straightforward to show that $\hat{p}_\mathrm{G}(\theta)$ is an unbiased estimator of $p(\theta)$, using the unbiasedness of $\hat{\gamma}$:
\begin{align}
    \mathbb{E} \left[ \hat{\gamma} (\mathcal{M}(X, \theta_\mathcal{L},\theta)) \right] &= \mathbb{E}_{\mathbb{Q}_{\theta_{\mathcal{L}}}} [ \mathcal{M}(X, \theta_\mathcal{L},\theta) \mathbf{1}_{\{\tau_B < \tau_A\}} ] \nonumber \\
    &= \mathbb{E}_{\mathbb{Q}_{\theta}} [ \mathbf{1}_{\{\tau_B < \tau_A\}} ] = p(\theta).
\end{align}
The use of the estimator in equation~\ref{p_g} enables us to rely exclusively on trajectories sampled at $\theta = \theta_\mathcal{L}$. However, in order to compute the Girsanov weights $\mathcal{M}(X, \theta_\mathcal{L}, \theta)$, it is necessary, for each point of the sampled trajectories, to evaluate the perturbed potential $V(x,\theta)$. This becomes computationally intractable when considering an entire distribution of parameters $\theta \sim \pi$. This limitation, coupled with the intractability of $\pi$ in high-dimensional nonlinear regimes (as noted in Section~\ref{sec:uq_linear}), justifies restricting the scope to linear potentials.

\subsection{Linear case}

Consider a potential belonging to the linear framework defined in section~\ref{sec:uq_linear}:
\begin{equation}
    V(\mathbf{x},\theta) = \theta^\top \mathcal{D}(\mathbf{x}).
\end{equation}
The associated drift is:
\begin{equation}
     -\nabla_\mathbf{x} V(\mathbf{x},\theta) = - [\nabla_\mathbf{x} \mathcal{D}(\mathbf{x})]^\top \theta.
\end{equation}
By equation \eqref{eq:likelihood_overdamped}, Girsanov reweighting factor for overdamped dynamics is:
\begin{equation}
    \mathcal{M}(X, \theta_\mathcal{L},\theta) = \frac{\exp[L(X, \theta_\mathcal{L},\theta)]}{Z(\theta_\mathcal{L},\theta)},
\end{equation}
with:
\begin{align} \label{eq:z_and_l}
    Z(\theta_\mathcal{L},\theta) &= {\mathbb{E}_{\lambda_{\partial{A}}^{\theta_{\mathcal{L}}}}
       [\exp[-\beta(\theta- \theta_\mathcal{L})^\top \mathcal{D}]]}, \\
        L(X,\theta_\mathcal{L},\theta)
&=
(\theta-\theta_\mathcal{L})^\top \mathbf{s}(X)
-
\frac{1}{2}
(\theta-\theta_\mathcal{L})^\top
\mathbf{I}(X)
(\theta-\theta_\mathcal{L}),
\end{align}
where the score $\mathbf{s}(X)$ and the Fisher Information Matrix (FIM) $\mathbf{I}(X)$ are defined as:
\begin{align}
    \mathbf{s}(X)
&=
-\beta \mathcal{D}(\mathbf{x}_0)
-
\sqrt{\frac{\beta}{2}}
\sum_{k=0}^{n_\tau-1}
[\nabla_\mathbf{x}\mathcal{D}(\mathbf{x}_k)]
\xi_k\sqrt{\Delta t},\\
    \mathbf{I}(X) &= \frac{\beta}{2} \sum _{k=0}^{n_\tau-1} [\nabla_\mathbf{x} \mathcal{D}(\mathbf{x}_k)][\nabla_\mathbf{x} \mathcal{D}(\mathbf{x}_k)]^\top \Delta t.
\end{align}

Thus, for linear potentials, computing $\hat{p}_\mathrm{G}$ reduces to evaluating the vector $\mathbf{s}$ and the matrix $\mathbf{I}$ along the sampled trajectories:

\begin{align} \label{eq:girsanov_linear}
& \forall \theta \in \mathbb{R}^P, \nonumber &\\ 
        & \hat{p}_\mathrm{G}(\theta) =  \frac{\hat{p}_{ \mathrm{ams}}(\theta_\mathcal{L})}{{N_TZ(\theta_\mathcal{L},\theta)}}
         \sum_{i=1}^{N_T}  \exp \big[(\theta -\theta_\mathcal{L})^\top \mathbf{s}(X) \nonumber \\ & - \frac{1}{2} (\theta -\theta_\mathcal{L})^\top \mathbf{I}(X) (\theta -\theta_\mathcal{L})\big] .
\end{align}

The computation of $Z$ is quite straightforward with $K$ samples $y_j \sim \lambda_{\partial{A}}^{\theta_\mathcal{L}}$.
\begin{equation}
    Z(\theta_\mathcal{L},\theta) \approx \frac 1 K \sum_{j=1}^K
    \exp\left[
    -\beta(\theta-\theta_\mathcal{L})^\top
    \mathcal D(\mathbf y_j)
    \right]
\end{equation}

This estimator, although unbiased and enabling the analytical propagation of uncertainty with respect to $\theta$, has several drawbacks. First, the computation of the FIM can be quite costly for true MLIPs. For Neural Network-based MLIPs such as \texttt{MACE}, calculating the FIM $\mathbf{I}$ requires evaluating the Jacobian $\nabla_\mathbf{x} \mathcal{D}$ along each trajectory, necessitating $P = 256$ automatic differentiation passes. It is worth noting that the score $\mathbf{s}$ only requires evaluating the term $[\nabla_\mathbf{x} \mathcal{D}(\mathbf{x}_t)] \mathrm{d}\mathbf{B}_t$, which is a Vector-Jacobian Product, and thus is significantly faster to compute. Second, this estimator can suffer from the statistical instability of Girsanov reweighting mentioned in Section~\ref{sec:theory}. This is why in practice, for real-world high-dimensional examples, the cumulant expansion~\cite{kendall1961advanced} is used, as it is known to stabilize heavy-tailed estimators~\cite{miao2014improved, singh2023variational} and has been previously applied in the context of thermodynamic reweighting~\cite{ceriotti2012inefficiency,imbalzano_uncertainty_2021}:
\begin{equation}
\langle \exp(L(X,\theta_\mathcal{L},\theta)) \rangle = \exp \left( \sum_{k=1}^{\infty} \frac{\kappa_k(L)}{k!} \right),
\end{equation}
where the values $\kappa_k(L)$ are the cumulants of the random variable $L(X,\theta_\mathcal{L},\theta)$. The series can be truncated as a function of the order of $(\theta-\theta_\mathcal{L})$. The first order is particularly interesting as it only requires the computation of the score:
\begin{equation}
    \hat{p}_\mathrm{C}(\theta) = \frac{\hat{p}_{\mathrm{ams}}(\theta_\mathcal{L})}{Z(\theta_\mathcal{L},\theta)} \exp \left( (\theta -\theta_\mathcal{L})^\top \bar{\mathbf{s}} \right),
\end{equation}
with $\bar{\mathbf{s}} = \frac{1}{N_T}\sum_{i=1}^{N_T} \mathbf{s}(X_i)$ the empirical mean of the score over the sampled trajectories. We expect this estimator to give a good idea of the variation of the rare-event probability, as it shares with $\hat{p}_G$ the first-order Taylor expansion, while being much easier to compute and achieving better numerical stability in high dimensions.

The expressions for both estimators are preserved in the underdamped case; the only modification occurs in the expressions for the score and the FIM, as detailed in Appendix~\ref{app:OBABO}.

\subsection{Practical implementation}
Analogously to the standard sampling estimators, we can aggregate $n_{\mathrm{real}}$ independent realizations of the path sampling algorithm to define the following empirical estimators. Given the computed committor probabilities $\{ \hat{p}^1_\mathrm{ams}(\theta_\mathcal{L}), \ldots, \hat{p}_{\mathrm{ams}}^{n_{\mathrm{real}}}(\theta_\mathcal{L})\}$ and the reactive trajectories $\{X_i ^j\}_{1 \le i \le N_T, 1\le j \le n_\mathrm{real}}$ generated by the path sampling algorithm, the Girsanov estimator and the Cumulant estimator can be expressed as
\begin{align}
    \hat{p}_\mathrm{G}(\theta)
    ={}&
    \frac{1}{n_{\mathrm{real}}}
    \sum_{j=1}^{n_{\mathrm{real}}}
    \left[
    \hat{p}^j_\mathrm{ams}(\theta_\mathcal{L})
    \cdot
    \frac{1}{N_T}
    \sum_{i=1}^{N_T}
    \mathcal{M}(X^j_i,\theta_\mathcal{L},\theta)
    \right],
\end{align}
\begin{align}
    \hat{p}_{\mathrm{C}}(\theta)
    ={}&
    \left(
    \frac{1}{n_{\mathrm{real}}}
    \sum_{j=1}^{n_{\mathrm{real}}}
    \frac{\hat{p}^j_\mathrm{ams}(\theta_\mathcal{L})}{Z(\theta_\mathcal{L},\theta)}
    \right)
    \exp\left[
    \left(
    \frac{1}{n_{\mathrm{real}}}
    \sum_{j=1}^{n_{\mathrm{real}}}
    \bar{\mathbf{s}}^j
    \right)^\top
    (\theta-\theta_\mathcal{L})
    \right].
\end{align}

The variance can be estimated using the empirical variance for $\hat{p}_\mathrm{G}$ and via the delta method for $\hat{p}_{\mathrm{C}}$. In order to reconstruct the full conditional distribution $p \mid \theta$ from these moments, an explicit distributional hypothesis is required. We chose a log-normal distribution, which naturally satisfies the non-negativity constraint of probabilities and is uniquely determined by its first two moments, though alternative choices could be considered. 
Ultimately, this framework provides a complete methodology to compute the conditional distribution $\mathbb{P}(p \mid \theta)$ solely using sampling data generated at the reference point $\theta = \theta_\mathcal{L}$.

The results of this section and the previous one can be readily integrated into a pipeline to estimate uncertainty-aware committor probabilities, as illustrated in Fig.~\ref{fig:schema}. Given a probability distribution $\pi(\mathrm{d}\theta)$ over the parameter space (or a relevant subset thereof) centered around the maximum likelihood estimator $\theta_\mathcal{L}$, the rare event algorithm is first evaluated at $\theta_\mathcal{L}$. Utilizing Girsanov reweighting, we can then derive estimators for the conditional distribution $p \mid \theta$. This allows for the direct propagation of parametric uncertainty into the probability space, ultimately yielding the full uncertainty-aware distribution $\mathbb{P}(p \mid \pi)$. The pseudo-code of our workflow is presented in Figure~\ref{fig:uq_pipeline}.
\begin{figure}[t]
\centering
\begin{algorithm}[H]
\caption{Uncertainty propagation pipeline}
\label{alg:uq_pipeline}
\begin{algorithmic}[1]

\REQUIRE Reference parameter $\theta_\mathcal{L}$, parameter distribution $\pi(d\theta)$

\STATE Run $n_{\mathrm{real}}$ independent AMS simulations at $\theta_\mathcal{L}$
\STATE Store the rare-event estimates and the reactive trajectories

\FOR{$k=1,\ldots,N_\theta$}
    \STATE Sample $\theta_k\sim\pi(d\theta)$
    \STATE Compute $\hat p_G(\theta_k)$ or $\hat p_C(\theta_k)$
    \STATE Fit a log-normal distribution to
    $\mathbb{E}[\hat p(\theta_k)]$ and $\mathrm{Var}[\hat p(\theta_k)]$
\ENDFOR

\STATE Approximate
\[
\mathbb{P}(p\mid\pi)
\approx
\frac{1}{N_\theta}
\sum_{k=1}^{N_\theta}
\mathbb{P}(p\mid\theta_k)
\]

\STATE \textbf{return} $\mathbb{P}(p\mid\pi)$
\end{algorithmic}
\end{algorithm}
\caption{Pseudo-code of the proposed uncertainty propagation framework.}
\label{fig:uq_pipeline}
\end{figure}

\section{Results}
\subsection{Parametric Sensitivity}\label{sec:sensitivity}
In this subsection, we present two toy models to validate the theoretical framework developed above. We perform a sensitivity analysis by varying the parameters of a reference potential to evaluate the accuracy of the estimators $\hat{p}_\mathrm{G}(\theta)$ and $\hat{p}_{\mathrm{C}}(\theta)$. The parameters studied correspond to linear parameterizations of the potentials, such as the height of the energetic barrier. The investigated potentials are the well-known solvated dimers and the rugged Müller–Brown potential, as illustrated in Figure~\ref{fig:potentials_toymodels}. All the parameters used in the simulations are listed in Appendix~\ref{app:simulation_parameters}.
\begin{figure}[htbp]
    \centering
    \begin{subfigure}[b]{0.49\textwidth}
        \centering
        \includegraphics[width=\textwidth]{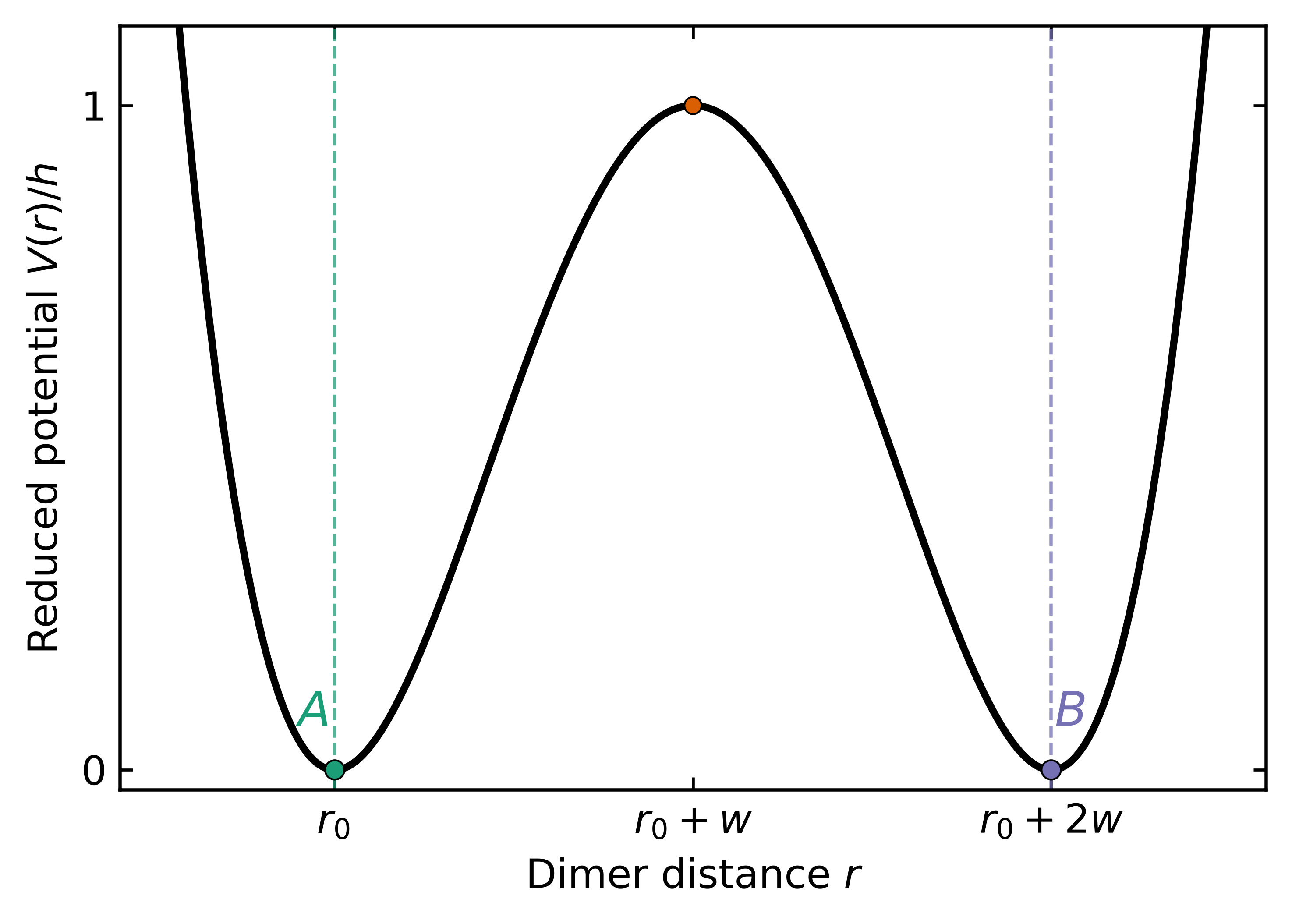}
        \label{fig:dimer_double_well}
    \end{subfigure}
    \hfill
    \begin{subfigure}[b]{0.49\textwidth}
        \centering
        \includegraphics[width=\textwidth]{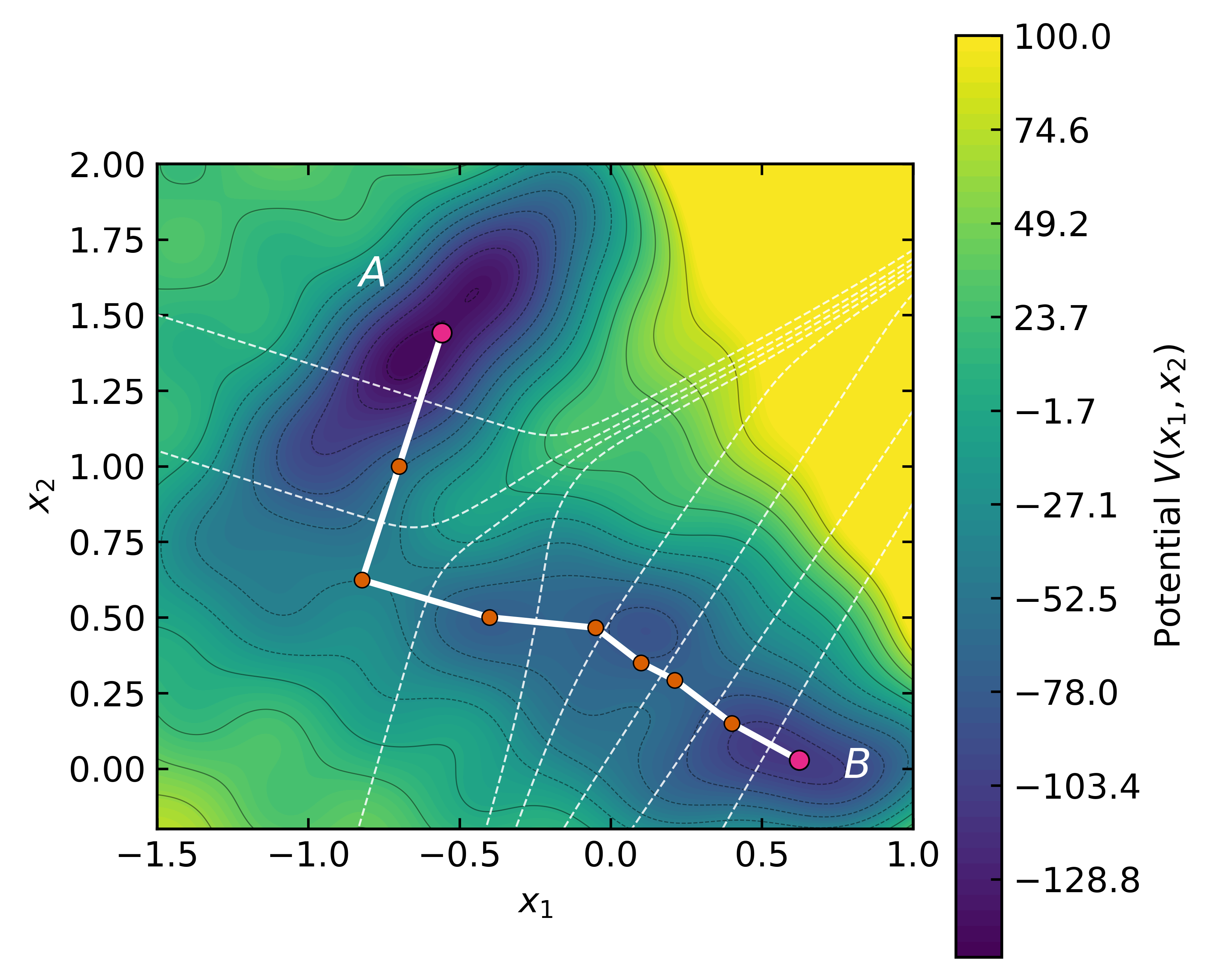}
        \label{fig:muller_brown}
    \end{subfigure}
    \caption{Left: Double-well potential modeling the interaction between the two dimer particles forming the dimer. Right: Rugged Müller-Brown potential, as well as the path collective variable used for the computations of AMS. The dashed lines represent isosurfaces of the collective variable.}
    \label{fig:potentials_toymodels}
\end{figure}

\subsubsection{Dimer in a WCA Solvent}

\begin{figure}[htbp]
    \centering
    \includegraphics[width=\linewidth]{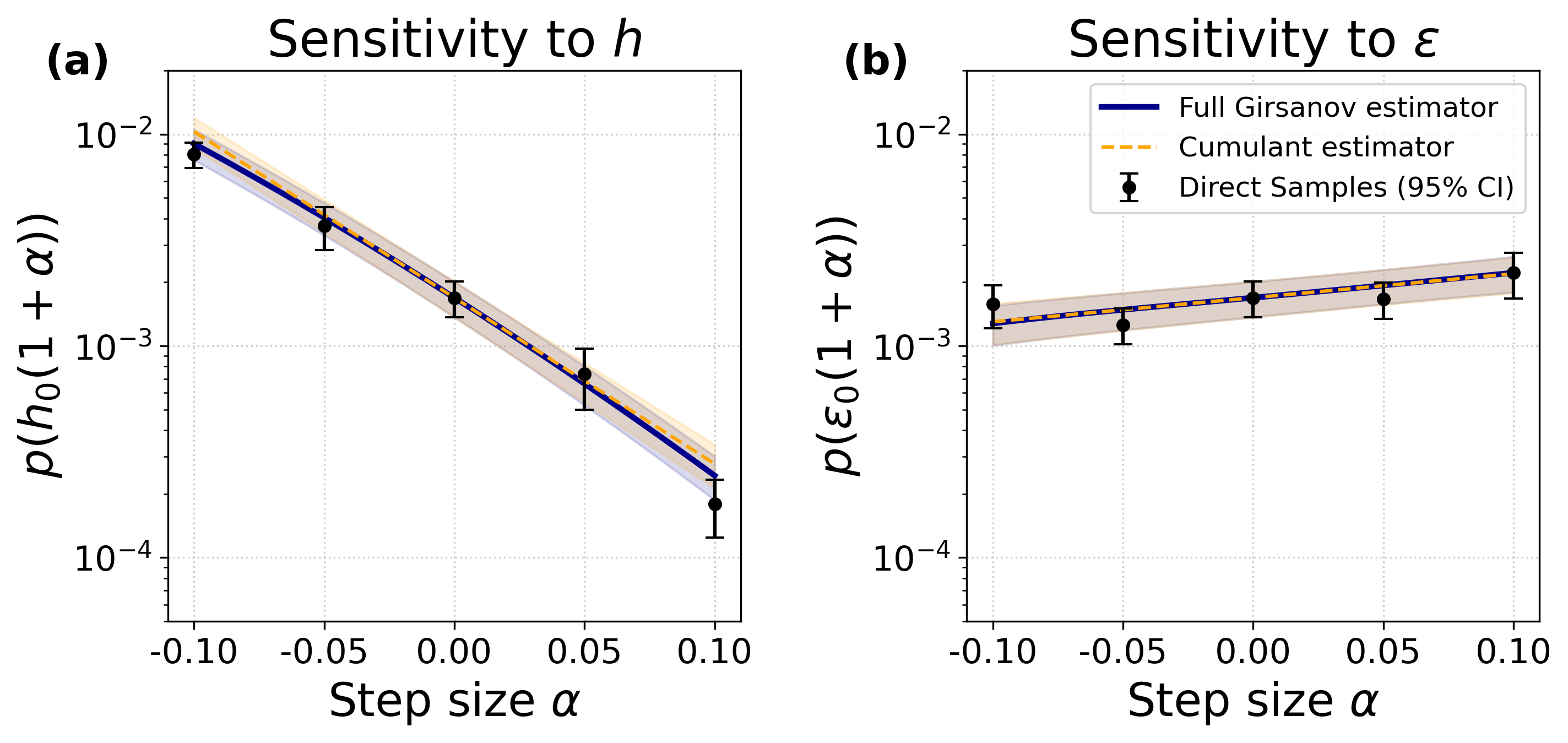}
    \caption{Sensitivity analysis of the dimer rare event probability (with $\theta_0 = (5,1)$) for $N=30$ particles. The variation is parameterized by $\alpha$ such that $h = h_0(1+\alpha)$ and $\epsilon = \epsilon_0(1+ \alpha)$. Direct samples, computed with AMS, are plotted in black with their corresponding confidence intervals. The estimators computed from the initial value are plotted in blue (estimator $\hat{p}_\mathrm{G}$) and orange ($\hat{p}_\mathrm{C}$) along with their confidence intervals.}
    \label{fig:dimer_SA}
\end{figure}

We consider a system of $N_\mathrm{WCA}$ particles in a three-dimensional periodic cell. Two of these particles form a dimer, while the remaining $N-2$ particles constitute the solvent. The solvent particles interact with each other and with the dimer via the WCA~\cite{weeks1971role} potential:
\begin{equation}
V_{\mathrm{WCA}}(r) = 4\epsilon \Bigl(\left[ \left(\frac{\sigma}{r}\right)^{12} - \left(\frac{\sigma}{r}\right)^6 \right]  \Bigl) \mathbf{1}_{\{r \le r_0\}}(r),
\end{equation}
where $r$ is the distance between two atoms, and $\epsilon, r_0$ and $\sigma$ are three positive parameters. The internal interaction between the two particles forming the dimer is governed by a double-well potential:
\begin{equation}
V_{\mathrm{D}}(r) = h \left[ 1 - \frac{(r - r_0 - w)^2}{w^2} \right]^2
\end{equation}
Consequently, the total potential energy of the system is given by:
\begin{align}
V(\mathbf{x}) = {}& V_{\mathrm{D}}(|x_1 - x_2|) + \sum_{3 \le i < j \le N_\mathrm{WCA}} V_{\mathrm{WCA}}(|x_i - x_j|) \nonumber \\
& + \sum_{i=1,2} \sum_{3 \le j \le N_\mathrm{WCA}} V_{\mathrm{WCA}}(|x_i - x_j|)
\end{align}
This total potential can be linearly parameterized with respect to the parameter vector $\theta = [h, \epsilon]^\top$ by introducing the descriptors $D_1(\mathbf{x})$ and $D_2(\mathbf{x})$:
\begin{equation}
V(\mathbf{x}) = \theta^\top \begin{bmatrix} D_1(\mathbf{x}) \\ D_2(\mathbf{x}) \end{bmatrix} = h D_1(\mathbf{x}) + \epsilon D_2(\mathbf{x})
\end{equation}
The potential energy surface exhibits two local minima. The first, denoted by $A = \{\mathbf{x} \mid |x_1-x_2| \le r_0\}$, corresponds to the compact state. The second, denoted by $B = \{\mathbf{x} \mid |x_1-x_2| \ge r_0 + 2w\}$, corresponds to the stretched state. The reaction coordinate used to describe the transition from the compact to the stretched state is the normalized bond length of the dimer molecule, denoted by $\xi$. We define the boundary
\[
\partial A=\{\mathbf{x}\mid |x_1-x_2|=r_0+\delta r\}.
\]
However, because this is only a low-dimensional toy model, we do not sample the corresponding exit distribution. Instead, all trajectories are initialized from a single configuration close to the reactant state $A$. Equivalently, $\lambda_{\partial A}$ is taken to be a Dirac measure centered at this configuration. We then compute the rare event probability
\[
p=\langle p_{A\to B}\rangle_{\lambda_{\partial A}},
\]
and perform a sensitivity analysis with respect to $\theta$ for $N=30$ particles, as shown in Fig.~\ref{fig:dimer_SA}.

As expected, the height of the energy barrier is the most sensitive parameter and is accurately captured by the Girsanov estimators. Since the response exhibits an exponential-like behavior, the first-order cumulant estimator $\hat{p}_\mathrm{C}$ achieves satisfactory coverage for substantial variations of the parameter $h$. It is worth emphasizing that, in practice, the investigated ranges of variation (between 90\% and 110\% of the nominal parameters) are significantly broader than what is typically required for standard uncertainty propagation applications.
\subsubsection{Rugged Müller-Brown Potential}

We consider a system described by a coordinate vector $\mathbf{x} = (x_1, x_2) \in \mathbb{R}^{2}$. The coordinates $(x_1, x_2)$ evolve on a rugged Müller-Brown potential surface.

The classical Müller–Brown surface is defined as a sum of four Gaussian terms:
\begin{align}
V_{\mathrm{MB}}(x_1, x_2) = \sum_{k=0}^{3} A_k \exp \Bigl[ & a_k(x_1 - \mathbf{x}_k)^2 + b_k(x_1 - \mathbf{x}_k)(x_2 - Y_k) \nonumber \\
& + c_k(x_2 - Y_k)^2 \Bigr]
\end{align}
A high-frequency periodic perturbation~\cite{e_transition-path_2010} is added to create local ruggedness:
\begin{equation}
V_{\mathrm{rug}}(x_1, x_2) = h_\mathrm{rug} \sin(\omega x_1) \sin(\omega x_2)
\end{equation}
The total energy of the system is therefore given by:
\begin{equation}
V(\mathbf{x}) = V_{\mathrm{MB}}(x_1, x_2) + V_{\mathrm{rug}}(x_1, x_2)
\end{equation}
We can parametrize this potential as a function of the Gaussian amplitudes $A_k$ and the perturbation amplitude $h_\mathrm{rug}$. 
\begin{equation}
V(\mathbf{x}) = \theta^\top D(\mathbf{x})
\end{equation}
where $\theta = [A_0, A_1, A_2, A_3, h_\mathrm{rug}]^\top \in \mathbb{R}^{5}$ and $D$ is the featurization associated with the Müller-Brown potential. 
The Müller-Brown potential exhibits three metastable states. Following the standard convention, the local minimum located at $	\approx (0.623, 0.028)$ is defined as the reactant state $A$, while the local minimum at $\approx (-0.558, 1.442)$ is defined as the product state $B$. The remaining minimum is considered an intermediate state $I$. Regarding the parameter values of the Müller-Brown potential, the amplitudes $A_0$, $A_1$, $A_2$, and $A_3$ correspond respectively to the depth of state $B$, the depth of the intermediate state ($I$), the depth of state $A$, and the height of a potential barrier located between the intermediate state and state $A$. 

\begin{figure}[htbp]
    \centering
    \includegraphics[width=\linewidth]{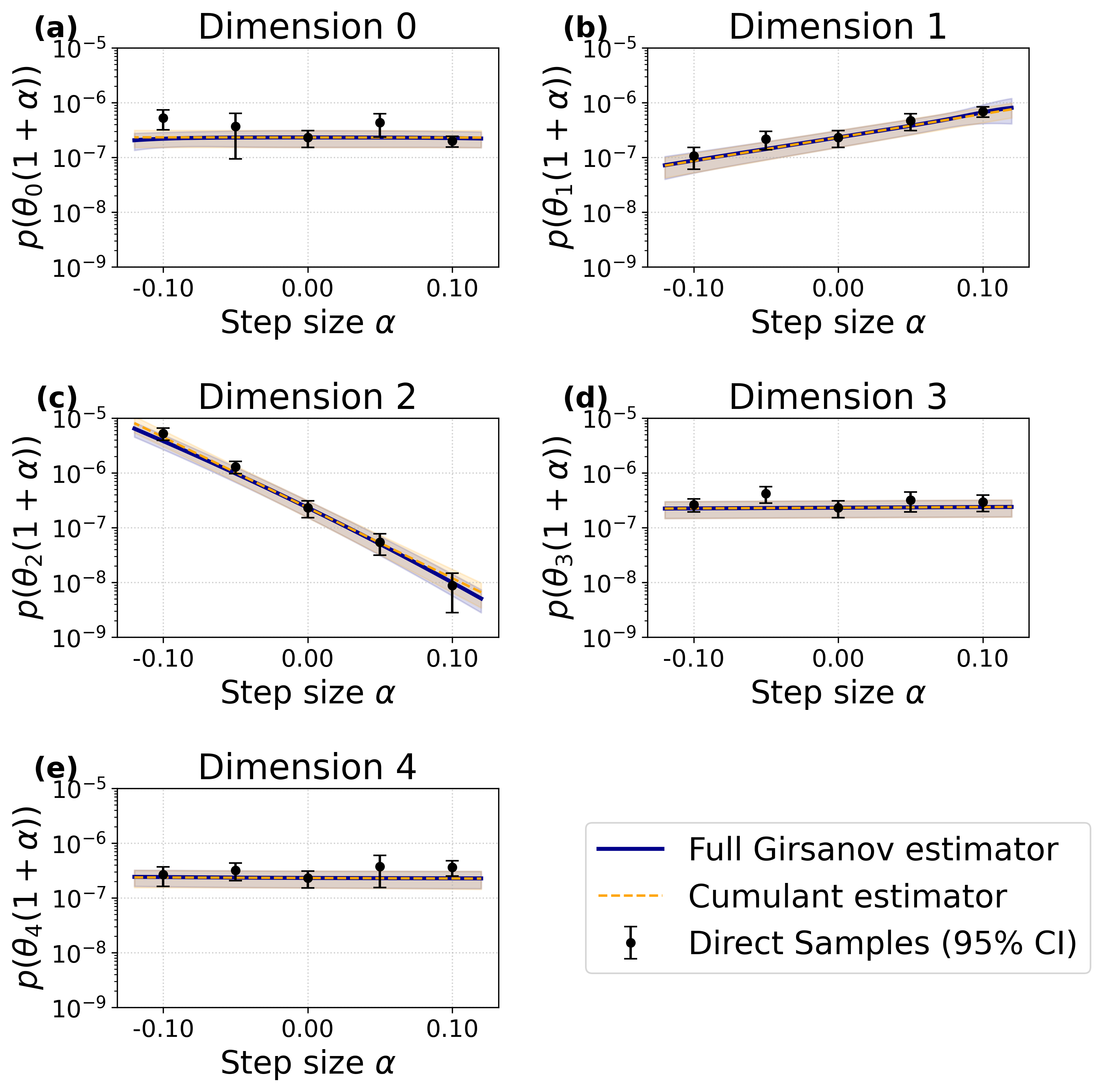}
    \caption{Sensitivity analysis of the rare event probability across all parameters around a reference parametrization $\theta_0 = [\theta_0^0, \dots, \theta_0^4]$. The variation is parameterized by $\alpha$ such that $\theta^i = \theta^i_0 (1+\alpha)$. Direct samples, computed with AMS, are plotted in black with their corresponding confidence intervals. The estimators computed from the initial value are plotted in blue (estimator  $\hat{p}_\mathrm{G}$)and orange ($\hat{p}_\mathrm{C}$) along with their confidence intervals.\\
    Dimension 0: depth of the $B$ state; Dimension 1: depth of the intermediate ($I$) state; Dimension 2: depth of the $A$ state; Dimension 3: height of the hill between the intermediate state and the $A$ state; Dimension 4: amplitude of the perturbations.}
    \label{fig:mb_SA}
\end{figure}

As for the dimer example, all trajectories are initialized from a single configuration close to the reactant minimum $\mathbf{x}_A$, corresponding to a Dirac measure $\lambda_{\partial A}$. We then compute the rare event probability $p=\langle p_{A\to B}\rangle_{\lambda_{\partial A}}$. We employ a path collective variable defined by a set of reference points close to the well-known minimum energy path, as illustrated in Figure~\ref{fig:potentials_toymodels}, since our objective is not to benchmark the AMS algorithm itself. We then perform a sensitivity analysis with respect to $\theta$, as shown in Figure~\ref{fig:mb_SA}.

Again, the results of the two estimators are very similar. Moreover, they match the direct AMS estimates very well for some parameters, even when the transition probability varies by several orders of magnitude. The most sensitive parameters correspond to dimensions 1 and 2. The dominant sensitivity is observed for parameters controlling the reactant and intermediate regions of the energy landscape, whereas perturbations affecting the product basin or local ruggedness have only a limited influence on the transition probability.

\subsection{Predictive Uncertainty Propagation}\label{sec:results_up}
In this section, we evaluate our uncertainty propagation pipeline on two distinct test systems. The first is a toy model involving a one-dimensional potential energy surface, which allows for a straightforward visualization of the potential's uncertainty. The second is a realistic physical system: the conformational space of a butane molecule, modeled using \texttt{MACE} foundation models~\cite{batatia2023foundation}.
\subsubsection{One-dimensional Potential}
\begin{figure}[htbp]
    \centering
    \includegraphics[width=\linewidth]{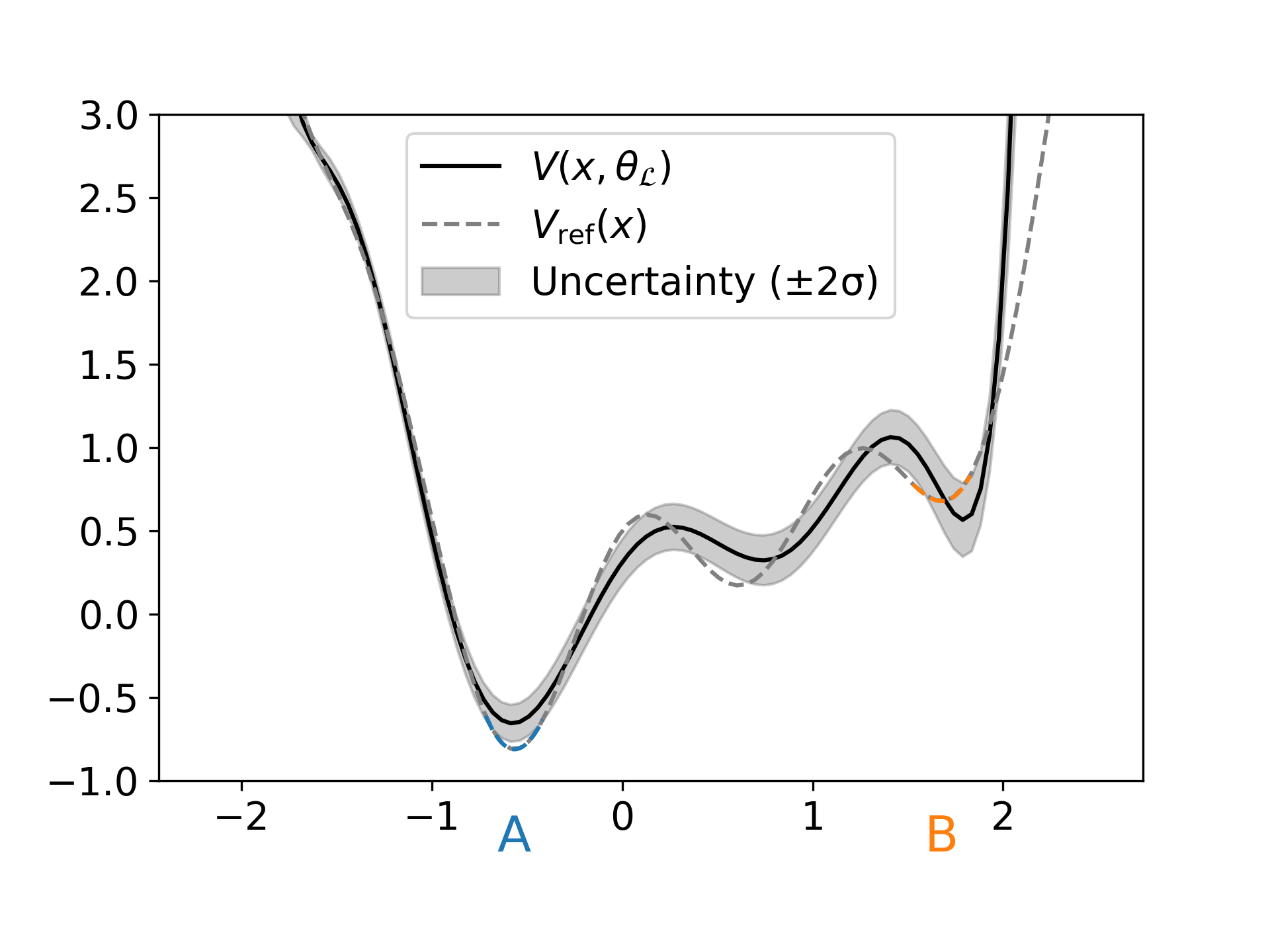}
    \caption{One-dimensional non-linear reference potential (dashed grey) fitted with an 8-degree polynomial (solid black). The parametric uncertainties computed via POPS are displayed as a shaded black area. The two metastable states $A$ and $B$ are defined by two narrow intervals around the local minima of the initial potential. }
    \label{fig:fittedpotential}
\end{figure}

We consider a one-dimensional reference potential $V_{\mathrm{ref}}$. We fit this potential using a POPS regression with a polynomial of degree 8. This optimization yields an MLE parameter vector $\theta_{\mathcal{L}} \in \mathbb{R}^9$ and a posterior distribution $\pi(\mathrm{d}\theta)$ over the parameter space, as illustrated in Fig.~\ref{fig:fittedpotential}.

\begin{figure}[htbp]
    \centering
    \includegraphics[width=\linewidth]{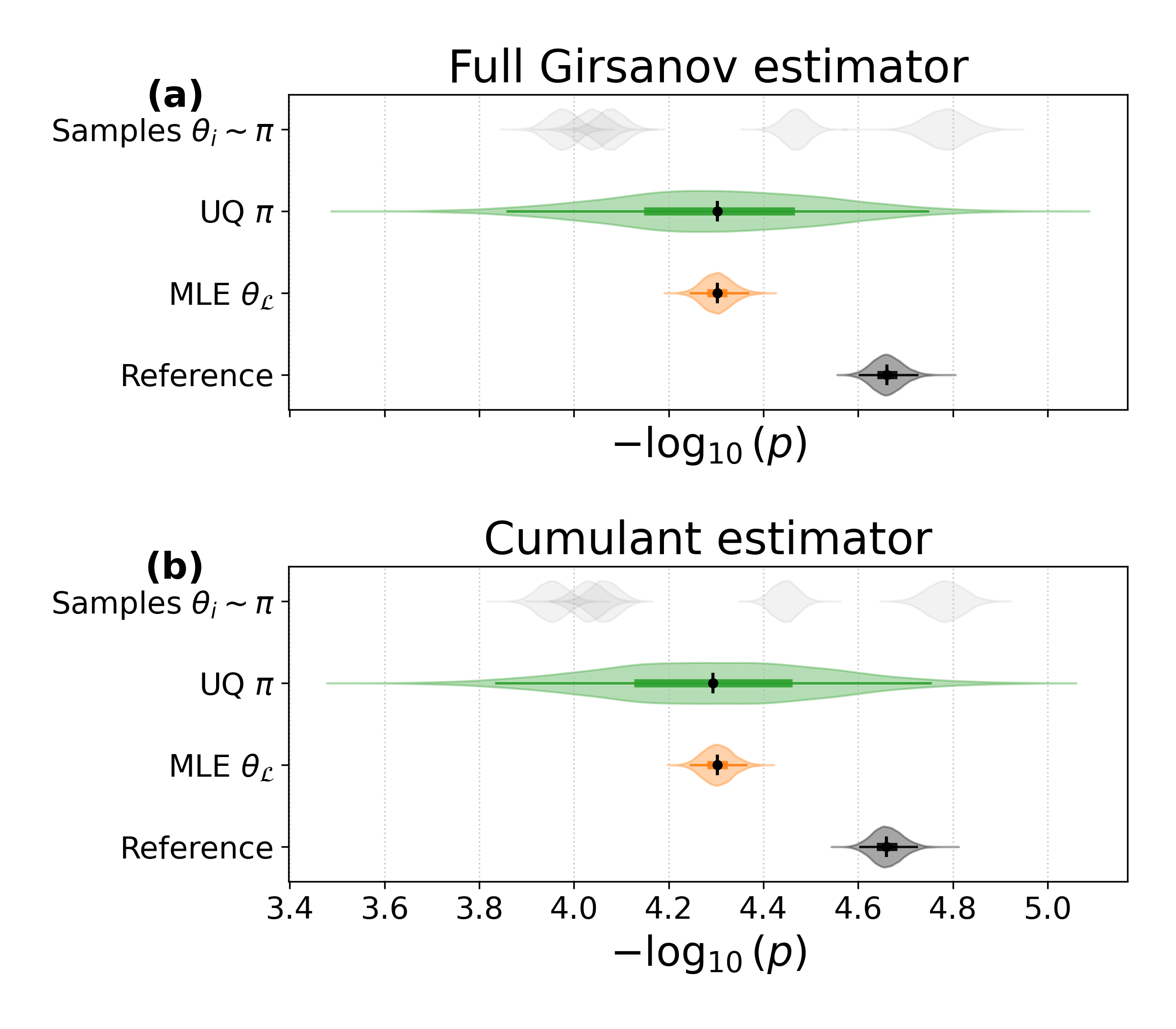}
    \caption{Violin plots of the probability density function of the rare event transition negative log probability on the one dimensional potential, using the full Girsanov estimator $\hat{p}_{\mathrm{G}}(\theta)$ (top) and cumulant approximation $\hat{p}_{\mathrm{C}}(\theta)$ (bottom). The reference distribution is shown in black, the prediction from the best-fit potential is in orange, and the full Bayesian uncertainty-aware distribution $p \mid \pi$ is in green. Samples of the parameter vector $\theta$ were drawn, and their corresponding conditional probability distributions are plotted in grey. }
    \label{fig:toy_model_combined}
\end{figure}

We compute the rare event probability (the transition from state $A$ to state $B$), using AMS, using both the reference and the fitted potentials. This yields the reference value $p_{\mathrm{ref}}$ and the nominal fitted distribution $\mathbb{P}(p \mid \theta_{\mathcal{L}})$. By applying the uncertainty propagation framework detailed in Section~\ref{sec:uncertainty_propagation} alongside our proposed estimators, we can evaluate the full propagated distribution $\mathbb{P}(p \mid \pi)$. 
The results obtained with the two Girsanov-based estimators $\hat{p} _ G$ and $\hat{p}_\mathrm{C}$ are reported in Fig.~\ref{fig:toy_model_combined} by plotting the distributions (reference, MLE, UQ) of the negative log probability $-\log p$. 
First, we see that the results are very similar for the two estimators. Note that the reference is itself a distribution, as it results from repeated AMS computations at the reference potential. We also see that the MLE prediction is quite far from the reference. In grey, we show the propagated distribution obtained from five parameter samples $\theta$ drawn according to $\pi$. These predictions are widely scattered. The UQ distribution $p \mid \pi$, obtained from the log-Gaussian mixture over the parameter samples, succeeds in encompassing the discrepancy between the reference and the MLE prediction while maintaining a reasonably sized confidence interval. This is exactly the type of uncertainty estimate we seek for the rare-event probability.

\subsubsection{Butane with MACE Foundation Models}\label{sec:butane}

Butane ($\mathrm{C}_4\mathrm{H}_{10}$) is a prototypical molecular system for studying conformational dynamics. Its conformational state is primarily characterized by the dihedral angle $\phi$ around the central $\mathrm{C}-\mathrm{C}$ bond. The underlying free energy surface exhibits three local minima: the global minimum at $\phi = 180^\circ$ (the \textit{trans} or \textit{anti} conformer) and two equivalent local minima located at $\phi \approx \pm 60^\circ$ (the \textit{gauche} conformers). 

We define the reactant state $A$ as the basin corresponding to the $\phi = 180^\circ$ minimum, and the product state $B$ as the union of the two basins around $\phi \approx \pm 60^\circ$. Utilizing Fleming-Viot particle systems~\cite{asselah2011quasistationary}, we generated a sample distributed according to the exit distribution on the reactant border $\partial A$. Our objective is to compute the rare event probability of hitting $B$ before returning to $A$, starting from $\partial A = \{\mathbf{x} \in \Omega,  \phi(\mathbf{x})\in \{175, 185\} \}$ at a temperature of $\mathcal{T} = 300$~K while providing a rigorous quantification of the uncertainty associated with this estimate.

Because performing a full DFT reference study for rare butane transitions would be computationally prohibitive, we use the \texttt{MACE-MPA-0 medium} model as a controlled reference potential. The objective of this section is therefore not to reproduce a realistic DFT-training workflow, but to assess whether the proposed uncertainty-propagation framework can recover the rare event probability of a known reference dynamics from imperfect surrogate MLIPs. Specifically, we designate the \texttt{MACE-MPA-0 medium} model as our reference “ground truth” and assess two reduced-size models, \texttt{MACE-OMAT-0 small} and \texttt{MACE-MP-0a small}, as surrogates. The \texttt{MACE-MP-0a} model represents the initial realization of the \texttt{MACE} foundation models and was trained on the \texttt{MPTraj} dataset~\cite{mptrj}, whereas \texttt{MACE-OMAT-0} was trained on the Open Materials dataset~\cite{omat}. The reference \texttt{MACE-MPA-0} model was trained on both \texttt{MPTraj} and \texttt{sAlex}~\cite{mptrj}, and was recently used for the determination of the committor probability in a complex catalytic reaction~\cite{pigeon_unbiased_2025}. By employing both distinct model capacities (small vs.\ medium) and training datasets, we aim to probe and quantify epistemic as well as model misspecification uncertainties.

\begin{figure}[htbp]
    
    \includegraphics[width = \linewidth]{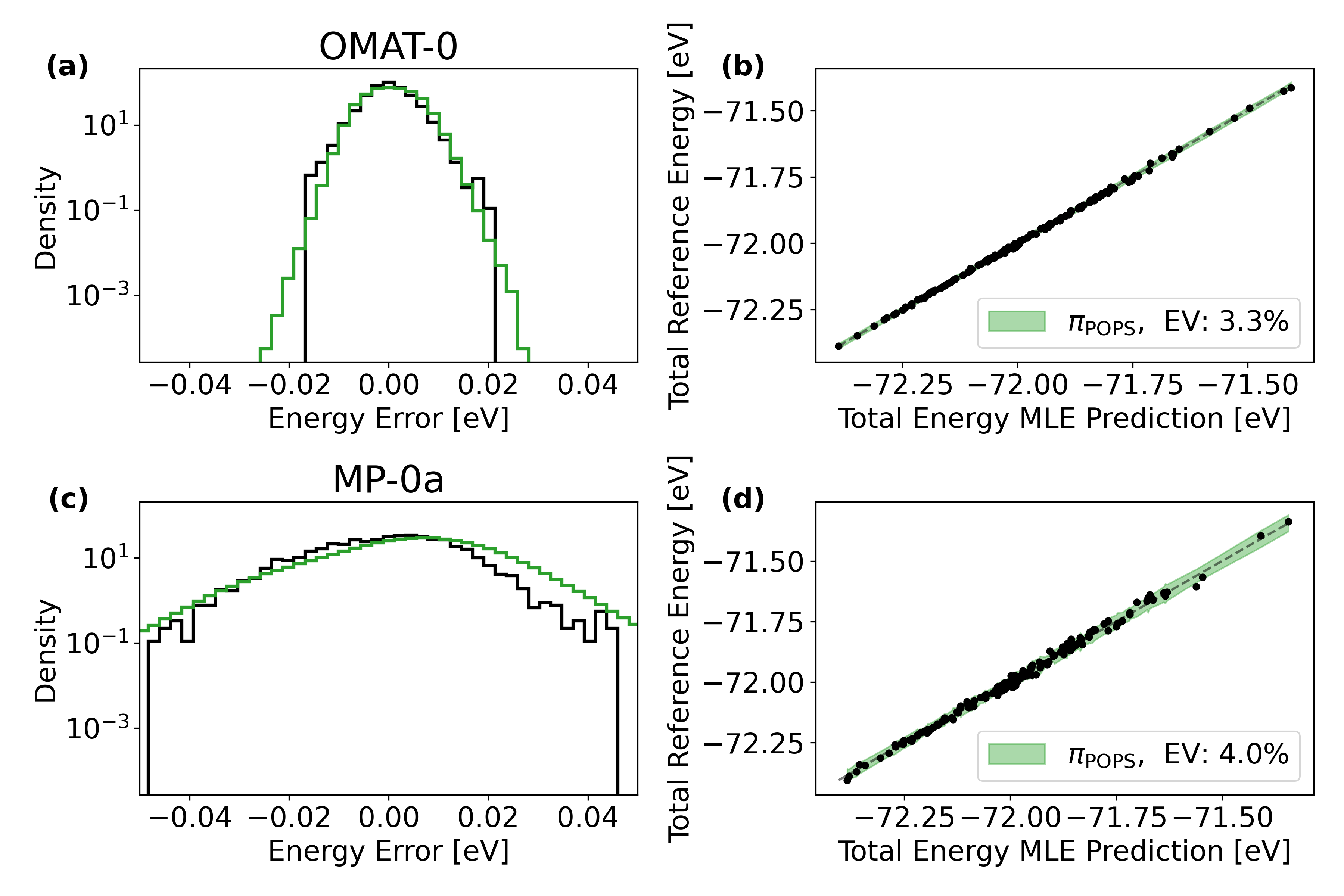}
\caption{Statistical analysis of local energy errors for the \texttt{MACE-OMAT-0} (top) and \texttt{MACE-MP-0a} (bottom) foundation models, from the posterior distribution $\pi_{\mathrm{POPS}}$. Left: Test error distributions for the MLE relative to the ground truth (black) and for the $\pi_{\mathrm{POPS}}$ distribution relative to the MLE (green). Right: Parity plot comparing reference energies against MLE predictions for a data subset. Shaded areas denote the 95 \% confidence interval of the predicted energy variable $y = V(x,\theta)$ where $\theta \sim \pi_{\mathrm{POPS}}$. The envelope violation (EV) measures the fraction of data points falling outside this confidence interval.}  \label{fig:uq_point_butane}
\end{figure}

The methodology is structured as follows:
\begin{enumerate}
    \item \textbf{Finetuning:} The two surrogate models are finetuned using $6400$ configurations sampled from the \texttt{MACE-MPA-0} reference potential by running trajectories starting in the metastable states $A$ and $B$.
    \item \textbf{Point-wise UQ:} POPS methodology is applied to the same configurations to retrieve a distribution $\pi(\mathrm{d}\theta)$ over the linear parameters of the surrogate model.
    \item \textbf{AMS computation:} AMS is computed at the MLE parameter $\theta_\mathcal{L}$ to retrieve the committor averaged probability $\hat{p}_{\mathrm{ams}}(\theta_\mathcal{L})$. All the parameters used in the simulations are referenced in Appendix~\ref{app:simulation_parameters}.
    \item \textbf{Propagation:} This parameter uncertainty is propagated to the rare event probabilities to characterize the distribution $\mathbb{P}(p \mid \pi)$. The cumulant estimator $\hat{p}_{\mathrm{C}}$ is used for faster computations and better \href{}{}numerical stability.
\end{enumerate} 
After finetuning, the two models achieve a Force Root Mean Square Error (FRMSE) of $30\,\text{meV/\AA}$ for \texttt{MACE-MP-0a} and $10\,\text{meV/\AA}$ for \texttt{MACE-OMAT-0}. We apply the POPS framework to the linear parameters of the MACE architecture to derive the parameter distribution $\pi_{\mathrm{POPS}}(\theta)$ around the MLE parameter $\theta_\mathcal{L}$. In practice, a sample of 2000 parameters was used. The resulting point-wise uncertainties are illustrated in Fig.~\ref{fig:uq_point_butane}. The two panels on the left of Fig.~\ref{fig:uq_point_butane} show the distribution of the true point-wise test errors and the error between the MLE prediction and the predictions of the UQ ensemble members. The close agreement between these distributions indicates that the resampled POPS ensemble $\pi_{\mathrm{POPS}}$ accurately reproduces the error statistics of the MLE.  The two right-hand panels of Fig.~\ref{fig:uq_point_butane} present the predictions and corresponding errors for a representative subset of the test set, together with their associated uncertainty bounds. Overall, the POPS framework exhibits excellent uncertainty quantification performance, as demonstrated by the strong agreement between the observed and predicted error distributions and by the Envelope Violation (EV) that is below $5\%$. As observed, \texttt{MACE-OMAT-0} achieves significantly higher accuracy compared to \texttt{MACE-MP-0a}.

\begin{figure}[htbp]
    \centering
    \includegraphics[width=\linewidth]{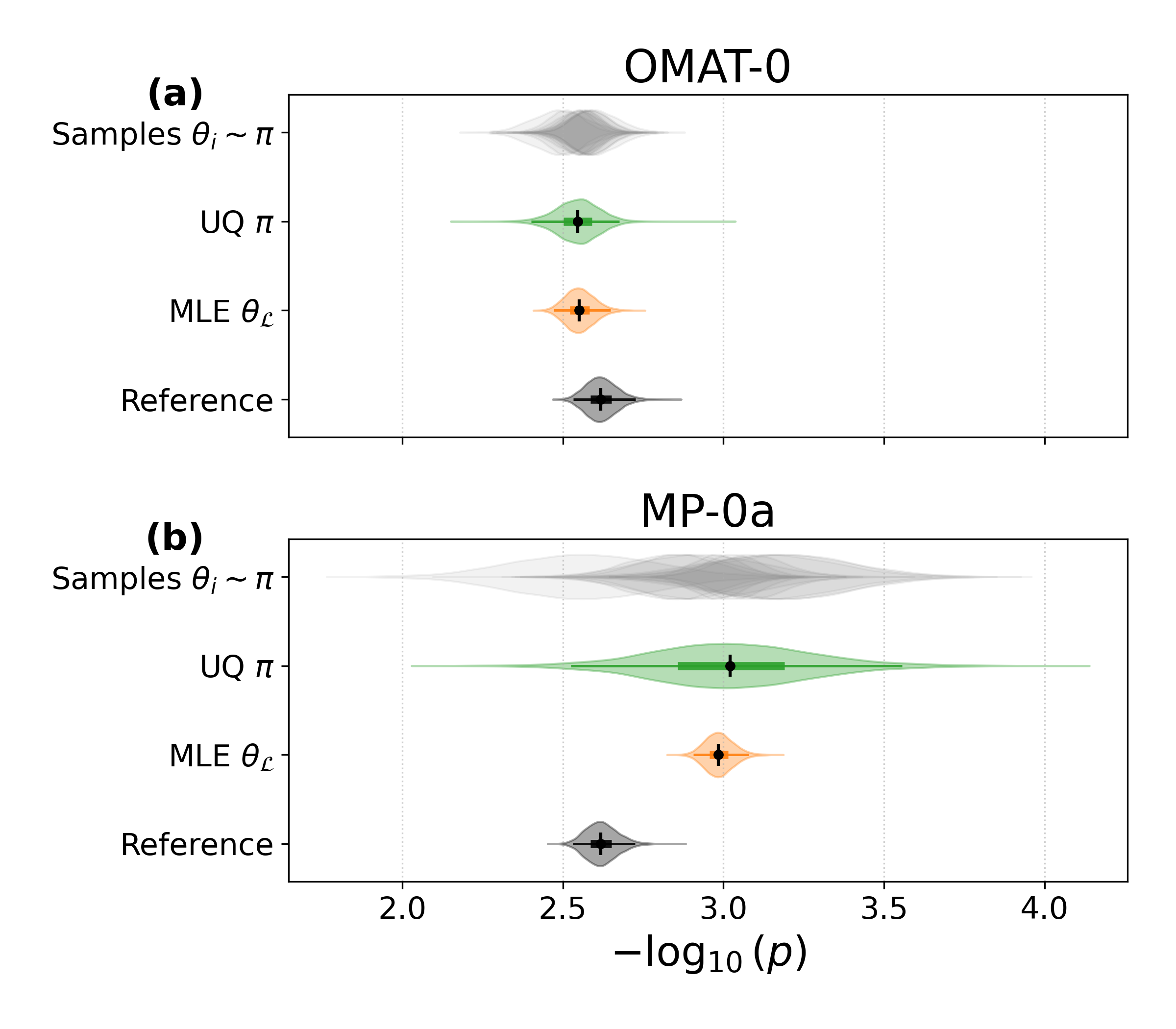}
    \caption{Violin plots of the probability density function of the negative log rare event probability of butane conformation, for \texttt{MACE-OMAT-0} (top) and \texttt{MACE-MP-0a} (bottom), at $\mathcal{T} = 300$~K. The reference distribution is shown in black, the prediction from the best-fit potential is in orange, and the full Bayesian uncertainty-aware distribution $p \mid \pi$ is in green. Samples of the parameter vector $\theta$ were drawn, and their corresponding conditional probabilities distributions are plotted in grey. }
    \label{fig:uq_prob_butane}
\end{figure}
We subsequently performed AMS at $\theta_{\mathcal{L}}$ (80 AMS runs with 100 replicas each) with both MLIPs and applied the uncertainty propagation framework detailed in the previous sections to compute the distribution and confidence intervals for the rare event committor probabilities. The results are presented in Fig.~\ref{fig:uq_prob_butane}. For the \texttt{OMAT} potential, the reference results and the ones at the MLE parameter $\theta_{\mathcal{L}}$ are nearly overlapping, with their respective Gaussians intersecting significantly. This consistency suggests that the model misspecification uncertainty is negligible compared to the epistemic (algorithmic) noise. Consequently, the full uncertainty-aware distribution $\mathbb{P}(p\mid \pi)$ remains centered near the original \texttt{OMAT} estimate, and the reference value is well-covered by the uncertainty bands. In contrast, for \texttt{MP-0a}, the distributions differ substantially from the reference. Here, the misspecification error due to the model is significantly more impactful than the epistemic noise. In this case, the uncertainty-aware density $\mathbb{P}(p \mid \pi)$ is notably wider, with its confidence interval successfully encompassing the reference \texttt{MPA} value. These results validate the effectiveness of our framework in capturing both types of uncertainty, even in cases of significant model bias.

The approach is also computationally efficient: a single 100-replica AMS run costs about one hour on GPU, and the reweighting step (score computation) adds roughly half an hour, after which inference for any subsequent parameter $\theta$ is essentially instantaneous. Compared to the direct approach of rerunning a full AMS simulation for each $\theta$, this yields a speedup of approximately $10^3\times$.

The computations were also conducted at temperature $\mathcal{T} = 200$~K and $\mathcal{T} = 500$~K. The results are quite similar and shown in Appendix~\ref{app:temperature}.


\subsection{Extension to the reaction rate}

The remaining step required to compute the full UQ of the reaction rate \(k_{AB}\) through Hill's relation is the quantification of uncertainty in \(\phi_A\). Indeed, given a distribution \(\mathbb{P}(\phi_A \mid \theta)\), uncertainty can be propagated according to
\begin{equation}
\begin{aligned}
\mathbb{P}(k_{AB} \mid \pi)
&=
\int_{\mathbb{R}^P}
\int_{0}^{+\infty}
\frac{1}{\phi_A}
\mathbb{P}\!\left(
p=\frac{k}{\phi_A}
\,\middle|\,
\theta
\right)
\\
&\qquad\qquad \times
\mathbb{P}\!\left(
\phi_A
\,\middle|\,
\theta
\right)
\, d\phi_A \,
\pi(d\theta).
\end{aligned}
\end{equation}
In practice, the method used to compute the exit frequency \(\phi_A\), both in the previous sections and in several earlier studies~\cite{lopes2019analysis,teo2016adaptive,pigeon_unbiased_2025}, consists of running long trajectories initiated in \(A\) and recording the points at which they cross \(\partial A\). Although a Girsanov reweighting could theoretically be applied, it would be computationally intractable due to the length of the trajectories.

Nevertheless, we argue that our method can still be applied to the reaction rate under certain assumptions. Our main observation is that MLIPs are often fine-tuned using data sampled from metastable states and are therefore substantially more accurate in these regions, as demonstrated in several active-learning studies~\cite{achar2025reactive,bachelor2025active,qi2024robust,kang2025accelerating}. This supports the hypothesis that the dominant source of uncertainty in Hill's relation  lies in the rare event contribution $p = \langle p_{A\to B}\rangle _{\lambda_{\partial{A}}}$, rather than in the exit frequency \(\phi_A\), which is estimated from trajectories sampled in the vicinity of \(A\). In our butane example, the exit-frequency estimates obtained with \texttt{MACE-OMAT-0} and \texttt{MACE-MP-0a} differed by less than \(3\%\) from the reference value computed using \texttt{MACE-MPA-0}, compared with discrepancies of approximately \(50\%\) in the rare event contribution, as shown in Table~\ref{tab:phi_a}.
\begin{table}[h]
\centering
\begin{tabular}{|l|c|}
\hline
Model & $\phi_A^{-1}$ (fs) \\
\hline
\texttt{MACE-OMAT-0 small} & 191.2 \\
\texttt{MACE-MP-0a small} & 186.5 \\
\texttt{MACE-MPA-0 medium} & 190.6 \\
\hline
\end{tabular}
\caption{Mean exit times from $A$ computed for the three models.}
\label{tab:phi_a}
\end{table}

Assuming that the estimation of \(\phi_A\) depends only weakly on \(\theta\), we can provide the uncertainty distribution of \(k_{AB}\) with respect to \(\pi\) with the equation:
\begin{equation}
\begin{aligned}
\mathbb{P}(k_{AB} \mid \pi)
&\approx
\int_{\mathbb{R}^P}
\int_{0}^{+\infty}
\frac{1}{\phi_A}
\mathbb{P}\!\left(
p=\frac{k}{\phi_A}
\,\middle|\,
\theta
\right)
\\
&\qquad\qquad \times
\mathbb{P}\!\left(
\phi_A
\,\middle|\,
\theta_{\mathcal L}
\right)
\, d\phi_A \,
\pi(d\theta).
\end{aligned}
\end{equation}
\begin{figure}[h!]
    \centering
    \includegraphics[width=1\linewidth]{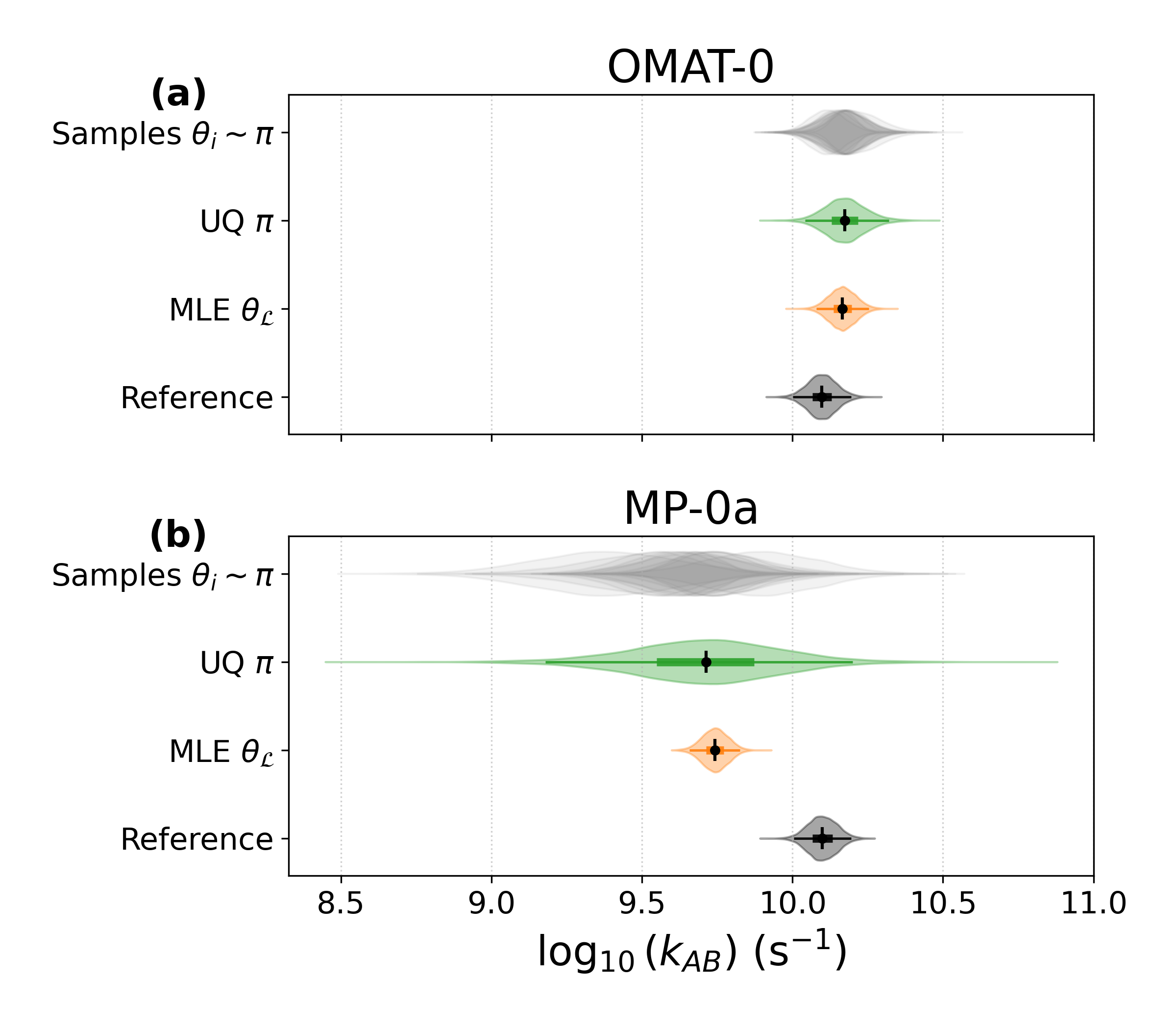}
    \caption{Violin plots of the probability density function of the reaction rate of butane conformation, for \texttt{MACE-OMAT-0} (top) and \texttt{MACE-MP-0a} (bottom). The reference distribution is shown in black, the prediction from the best-fit potential is in orange, and the full Bayesian uncertainty-aware distribution $k_{AB} \mid \pi$ is in green. Samples of the parameter vector $\theta$ were drawn, and their corresponding conditional probabilities distributions are plotted in grey.}
    \label{fig:uq-k}

\end{figure}
The distribution of the exit frequency \(\mathbb{P}\left( \phi_A \;\middle|\; \theta_{\mathcal{L}}\right)\) can therefore be estimated by sampling trajectories starting from $A$ at \(\theta=\theta_{\mathcal{L}}\). The results obtained under this assumption are shown in Fig.~\ref{fig:uq-k}. This framework can therefore be applied in the present setting, as it yields results similar to those obtained for the averaged committor. In particular, it provides a meaningful bound on the error while accurately capturing its magnitude.

\section{Conclusion and discussion}\label{sec:conclusion}

In this work, we presented a new methodology to quantify and propagate uncertainties in rare event kinetics driven by MLIPs. By coupling interacting particle systems, such as Adaptive Multilevel Splitting, with Girsanov path reweighting, we demonstrated that the parametric uncertainty inherent to misspecified MLIPs can be rigorously mapped onto macroscopic kinetic observables, such as the rare event probability of transition between two metastable states $p = \langle p_{A\to B}\rangle_{\lambda_{\partial{A}}}$.
A central contribution of this study is the derivation of computationally tractable Girsanov-based estimators. We showed that the first-order cumulant estimator, $\hat{p}_{\mathrm{C}}(\theta)$, achieves high accuracy while bypassing the prohibitive cost of computing the full Fisher Information Matrix along trajectories. This formulation is particularly advantageous for deep learning architectures where parameter spaces are vast. Our method was validated on several benchmark systems: a rugged Müller-Brown potential, a dimer potential in a solvent, and the conformational transitions of butane with MACE foundation potentials, demonstrating the stability and accuracy of our approach. In particular, the butane conformational application shows that this framework successfully isolates algorithmic variance from model misspecification, providing tight and reliable confidence intervals for the calculated rare event probabilities. Finally, we demonstrated that this uncertainty quantification on the rare event probability provides a very accurate approximation of the uncertainty quantification on the transition rate $k_{AB}$, since MLIPs are generally highly precise in the vicinity of the metastable states.

This last result is based on an assumption regarding the accuracy of MLIPs near the metastable basin and should be investigated in future work. Indeed, the choice of $\partial A$ is user-dependent and may, in more complex reactions, lie beyond the safe region where MLIPs are highly accurate. A method to propagate uncertainty to $\phi_A$, which does not require to apply Girsanov reweighting to potentially long trajectories, remains to be found. Another perspective is the application of our method to more complex activated processes, such as catalytic reactions. The goal would be to verify whether the variance remains tractable for longer and higher-dimensional trajectories. Finally, this Girsanov reweighting framework could pave the way for direct path-space observable reweighting from MLIPs to the ab initio DFT level.
\paragraph{Acknowledgments}{The authors kindly thank Tony Lelièvre for fruitful discussions leading to completion of this work.}



\paragraph{Data availability statement}{
The final processed datasets used to generate the figures, together with the scripts for data generation and analysis, as well as an explanatory notebook, are available at: \url{https://github.com/21moracchi/girsanov-uq}.}

\appendix
\section{Extension to underdamped Langevin} \label{app:OBABO}
In the underdamped regime, the definitions are slightly modified, as the dynamical path now includes both positions and momenta: 
\[
X = (\mathbf{x}_0,\mathbf{p}_0,\ldots, \mathbf{x}_{n_\tau},\mathbf{p}_{n_\tau}).
\]
Accordingly, the committor function \(p_{A \to B}\) is defined on the full phase space of positions and momenta, and the distribution \(\lambda_{\partial{A}}\) is evaluated on phase-space points located on \(\partial A\) whose velocities are directed outward from \(A\):
\begin{align}
    \lambda_{\partial{A}}(\mathbf{x},\mathbf{p})
    \propto {}&
    e^{-\beta H(\mathbf{x}, \mathbf{p})}
    \left[(M^{-1}\mathbf{p})^T \mathbf{n}_A(\mathbf{x})\right]_+
    \nonumber\\
    & \times
    \mathbf{1}_{\{(M^{-1}\mathbf{p})^T \mathbf{n}_A(\mathbf{x}) > 0\}}
    \,\sigma_{\partial A}(d\mathbf{x})\, d\mathbf{p},
\end{align}
where \(\sigma_{\partial A}\) denotes the surface measure on \(\partial A\), \(M\) is the mass matrix, and \(\mathbf{n}_A(\mathbf{x})\) is the outward unit normal vector to \(A\) at the point \(\mathbf{x} \in \partial A\).

For Girsanov reweighting, the derivation is more involved than the overdamped case~\cite{kieninger_girsanov_2023}. Since modern geometric "splitting" integrators treat the drift and Ornstein-Uhlenbeck operators separately, the exact expression for Girsanov weights depends on the specific integration scheme. Furthermore, the existence of a discrete-time path measure is not guaranteed for all schemes. In this work, we employ the OBABO scheme, for which the Girsanov weight is well-defined~\cite{kieninger_girsanov_2023}. 
The discrete-time update from step $n$ to $n+1$ via the quasi-symplectic \textbf{OBABO} splitting pattern is defined by the following sequential system of equations:

\begin{align}
\mathbf{O} \quad & \mathbf{p}^{n+1/4} = \frac{1}{c_1} \left( c_2 \, \mathbf{p}^n + \mathbf{C}_3 \, \xi^n \right) \\
\mathbf{B} \quad & \mathbf{p}^{n+1/2} = \mathbf{p}^{n+1/4} + \frac{\Delta t}{2} F(\mathbf{x}^n) \\
\mathbf{A} \quad & \mathbf{x}^{n+1} = \mathbf{x}^n + \Delta t \, M^{-1} \mathbf{p}^{n+1/2} \\
\mathbf{B} \quad & \mathbf{p}^{n+3/4} = \mathbf{p}^{n+1/2} + \frac{\Delta t}{2} F(\mathbf{x}^{n+1}) \\
\mathbf{O} \quad & \mathbf{p}^{n+1} = \frac{1}{c_1} \left( c_2 \, \mathbf{p}^{n+3/4} + \mathbf{C}_3 \, \eta^n \right)
\end{align}
with $F = -\nabla V$, $\xi^n, \eta^n \sim \mathcal{N}(0,\mathbf{I})$ independent Gaussian noises in $\mathbb{R} ^{3N}$ and $c_1,c_2$ scalars, and $\mathbf{C}_3$ a matrix defined by:
\begin{equation}
c_1 = 1 + \frac{\zeta \Delta t}{4}
\end{equation}

\begin{equation}
c_2 = 1 - \frac{\zeta \Delta t}{4}
\end{equation}

\begin{equation}
\mathbf{C}_3 = \sqrt{\frac{\zeta \Delta t}{\beta}} M^{1/2} = \sqrt{\zeta \Delta t \, k_B T} M^{1/2}
\end{equation}

As derived by Kieninger et al.~\cite{kieninger_girsanov_2023}, the Girsanov likelihood ratio for potentials $V$ and $\tilde{V} = V + U$ reads:
\begin{equation}
\begin{split}
    \frac{\mathrm{d}\mathbb{Q}_{\tilde{V}}}{\mathrm{d}\mathbb{Q}_V} (X) = \frac{\mu_{\tilde{V}}(\mathbf{x}_0)}{\mu_{V}(\mathbf{x}_0)}\exp \Big( -\sum_{n=0}^{n_\tau-1} \Big[ &\eta_n \cdot \Delta \eta_n + \xi_n \cdot \Delta \xi_n \\
    &+ \frac{1}{2} \left( \|\Delta \eta_n\|^2 + \|\Delta \xi_n\|^2 \right) \Big] \Big)
\end{split}
\end{equation}
where $\mu_{\tilde{V}}$ and $\mu_V$ refer to the probability measure on the initial point induced by the potentials, and with
\begin{equation}
    \Delta \xi_n = \frac{c_1 \Delta t}{2} \mathbf{C}_3^{-1} \nabla U(\mathbf{x}^n)
\end{equation}
\begin{equation}
    \Delta \eta_n = \frac{c_2 \Delta t}{2} \mathbf{C}_3^{-1} \nabla U(\mathbf{x}^{n+1})
\end{equation}

For a general linear potential $V(\mathbf{x},\theta) = \theta^\top \mathcal{D}(\mathbf{x})$, we can derive the path-dependent score vector $\mathbf{s}(X) \in \mathbb{R}^P$ and Fisher Information Matrix $\mathbf{I}(X) \in \mathbb{R}^{P \times P}$:

\begin{equation}
    \mathbf{s}(X) = -\beta\mathcal{D}(\mathbf{x}_0) - \frac{\sqrt{\beta \Delta t}}{2\sqrt{\zeta}} \sum_{n=0}^{n_\tau-1} \left[ c_1 [\nabla_\mathbf{x} \mathcal{D}(\mathbf{x}^n)] M^{-1/2} \, \xi^n + c_2 [\nabla_\mathbf{x} \mathcal{D}(\mathbf{x}^{n+1})] M^{-1/2} \, \eta^n \right]
\end{equation}

\begin{equation}
\mathbf{I}(X) = \frac{\beta \Delta t}{4\zeta} \sum_{n=0}^{n_\tau-1} \left[ c_1^2 [\nabla_\mathbf{x} \mathcal{D}(\mathbf{x}^{n})] M^{-1} [\nabla_\mathbf{x} \mathcal{D}(\mathbf{x}^{n})]^\top + c_2^2 [\nabla_\mathbf{x} \mathcal{D}(\mathbf{x}^{n+1})] M^{-1} [\nabla_\mathbf{x} \mathcal{D}(\mathbf{x}^{n+1})]^\top \right]
\end{equation}

The expression of $Z(\theta_\mathcal{L},\theta)$ remains identical to the one in Equation~\ref{eq:z_and_l}.
\section{Additional details on Adaptive Multilevel Splitting}\label{app:ams}

The Adaptive Multilevel Splitting (AMS) algorithm is designed to reduce the variance of standard Monte Carlo simulations when sampling rare events. 

Let $\xi: \Omega \to \mathbb{R}$ be a reaction coordinate. We define the reactive state $A$  and the product state $B$  such that
\begin{align}
A &= \{ \mathbf{x} \in \mathbb{R}^d \mid \xi(\mathbf{x}) < L_0 \}, \\
B &= \{ \mathbf{x} \in \mathbb{R}^d \mid \xi(\mathbf{x}) \ge L_{max} \}.
\end{align}
with two scalar values $L_0$ and $L_{max}$.
For any subset $V \subset \mathbb{R}^d$, let $\tau_V(\mathbf{x}) = \inf\{ t > 0 \mid q_t \in V \}$ denote the first entrance time of a stochastic process $q_t$ into $V$. We also define the global extinction/absorption time $\tau$ as
\begin{equation}
    \tau = \min\bigl(\tau_A, \tau_B\bigr).
\end{equation}
The quantity of interest is the rare event probability $p = \mathbb{P}^{x_0}\bigl(\tau_B < \tau_A \bigr)$.

In standard fixed splitting methods, a deterministic sequence of $M$ intermediate levels $L_0 < L_1 < \dots < L_M = L_{max}$ is introduced to factorize the probability via the nested formula
\begin{equation}\label{eq:splitting_equation}
    p = \prod_{j=0}^{M-1} \mathbb{P}^{x_0}\bigl( \tau_{\Sigma_{j+1}} < \tau_A \bigm| \tau_{\Sigma_j} < \tau_A \bigr),
\end{equation}
where $\Sigma_j = \{ \mathbf{x} \in \mathbb{R}^d \mid \xi(\mathbf{x}) = L_j \}$.

Conversely, the distinctive feature of the AMS method~\cite{cerou2019adaptive} is that the intermediate levels are determined adaptively online to minimize the variance. Starting with $N$ replicas, at each iteration $j$, the next level $L_{j+1}$ is defined as the score value of the $K$-th least advanced replica. The detailed pseudo-code of the algorithm is provided in Fig.~\ref{fig:ams_algo}. In this work, the $K=1$ variant of the algorithm is utilized~\cite{charles-edouard_unbiasedness_2015,cerou2019adaptive}, as it is known to minimize the asymptotic variance.

\begin{figure}[t]
    \centering
\begin{algorithm}[H]
\caption{Adaptive Multilevel Splitting (AMS) }
\label{alg:ams}
\begin{algorithmic}[1]
\REQUIRE Initial distribution $\lambda_{\partial{A}}$, importance function $\xi$, 
number of replicas $N_T$, minimal number $K$ of trajectories to discard at each iteration
\STATE \textbf{Initialization:}
\STATE Sample $\mathbf{x}_0^1, \dots, \mathbf{x}_0^{N_T} \sim \lambda_{\partial{A}}$ i.i.d.
\STATE Set $j \leftarrow 0$
\FOR{$i = 1$ to $N_T$}
    \STATE Run trajectory $X^i_0$ until its final time $\tau_i$
    \STATE Set $z_i \leftarrow \max\mathbf{x}_{0 \le t \le \tau_i} \xi(\mathbf{x}_{t}^i)$
\ENDFOR
\STATE Sort the $z_i$ in increasing order: $z_{(1)} \le \cdots \le z_{(N_T)}$
\STATE Set current level $L \leftarrow z_{(K)}$
\WHILE{$L < L_\mathrm{max}$}
    \STATE $j \leftarrow j + 1$
    \STATE Discard all trajectories such that $z_i \le L$
    \STATE Let $K_j$ be the number of discarded trajectories (with $K_j \ge K$)
    \STATE Let $I_j$ be the index set of the remaining trajectories
    \FOR{$i \in \{1,\dots,N_T\} \setminus I_j$}
        \STATE Choose uniformly at random an index in $I_j$
        \STATE Clone the corresponding trajectory until the first time $t$ such that $\xi(\mathbf{x}_t) > L$
        \STATE From that time, simulate the cloned trajectory until its final time $\tau_i$
        \STATE Replace trajectory $i$ by the new one
        \STATE Set $z_i \leftarrow \max\mathbf{x}_{0 \le t \le \tau_i} \xi(\mathbf{x}_{t}^i)$
    \ENDFOR
    \STATE Sort the $z_i$ in increasing order: $z_{(1)} \le \cdots \le z_{(N)}$
    \STATE Set current level $L \leftarrow z_{(K)}$
\ENDWHILE
\STATE Set $M = j$ 
\STATE \textbf{Estimate the rare event probability:}
\STATE
\[
\hat{p}_{\text{AMS}}
= \left( \prod_{j=1}^{M-1} \frac{N_T - K_j}{N_T} \right)
\]
\end{algorithmic}
\end{algorithm}
\caption{Pseudo-code of the AMS algorithm}
    \label{fig:ams_algo}
    
\end{figure}
\section{Bayesian Regression and POPS}\label{app:POPS}

For linear regression, the standard assumption on the data-generating process relies on a noisy linear response:
\begin{equation}
y(\mathbf{x}) = \theta^T \Phi(\mathbf{x}) + \epsilon, \quad \epsilon \sim \mathcal{N}(0, \sigma^2),
\end{equation}
where $y(\mathbf{x})$ is the ground-truth observation, $\Phi:\Omega \to \mathbb{R}^P$ is a feature map, and $\sigma \in \mathbb{R}_+^*$ is a positive scalar.   
Assuming a Gaussian prior $\pi_0(\theta) = \mathcal{N}(\mu_0, \Sigma_0)$, the analytical posterior distribution is given by:
\begin{equation}
\pi_{\mathrm{bayesian}}(\theta) = \mathcal{N}(\mu_n, \Sigma_n),
\end{equation}
where the updated mean and covariance matrix are defined as:
\begin{align}
\Sigma_n &= \left( \Sigma_0^{-1} + \frac{1}{\sigma^2} \Phi_\mathcal{X}^T \Phi_\mathcal{X} \right)^{-1}, \\
\mu_n &= \Sigma_n \left( \Sigma_0^{-1} \mu_0 + \frac{1}{\sigma^2} \Phi_\mathcal{X}^T \mathbf{y}_\mathcal{X} \right),
\end{align}
where $\Phi_\mathcal{X}$ and $\mathbf{y}_\mathcal{X}$ denote the design matrix and the label vector of the training dataset $\mathcal{X}$, respectively. 
In the case of a vanishing aleatoric error and over-parameterization, the posterior $\pi_{\mathrm{bayesian}}$ converges to a Dirac distribution centered on the optimal parameter $\theta_\mathcal{L}$, leading to a divergent generalization error:
\begin{equation}
\lim_{n/P \to + \infty} \mathcal{G}[\pi_{\mathrm{bayesian}}] = \mathcal{O}(1/\sigma ^2).
\end{equation}

The POPS method was introduced to mitigate this divergence. In brief, for each point $\mathbf{x}$ in the training dataset $\mathcal{X}$, it defines the point-wise exact-match manifold:
\begin{equation}
\mathcal{P}(\mathbf{x}) = \{\theta \mid f(\mathbf{x},\theta) = y(\mathbf{x})\},
\end{equation}
where $f: \Omega \times \mathbb{R}^P \to \mathbb{R}$ is the predictive model. 
It has been shown that the distribution minimizing the generalization loss $\mathcal{G}$ must assign weights within every POPS manifold. 
The per-point optimal parameter vector $\theta_\mathbf{x}^*$ is then formulated as the solution to the constrained optimization problem:
\begin{equation}
\theta_\mathbf{x}^* = \arg\min_{\theta \in \mathcal{P}(\mathbf{x})} \mathcal{L}(\mathbf{x}, \theta),
\end{equation}
where $\mathcal{L}(\mathbf{x}, \theta)$ represents the global loss function over the entire dataset $\mathcal{X}$. In the case of a linear model $f(\mathbf{x},\theta) = \theta ^\top \Phi(\mathbf{x})$, an analytic form of $\theta_\mathbf{x}^*$ can be derived. A uniform empirical distribution is subsequently constructed either directly over the set of local optima $\{\theta_\mathbf{x}^*\}$ or over the bounding hypercube enclosing these parameters, yielding the POPS parameter distribution $\pi_{\mathrm{POPS}}(\theta)$.

\section{Simulation Parameters}\label{app:simulation_parameters}

\begin{table}[H]
\caption{Algorithmic and physical parameters for the toy models. Units for the Dimer in WCA are dimensionless (Lennard-Jones units). Units for the Müller-Brown and 1D potential are standard physical units ($\mathrm{fs}$, $\mathrm{K}$, $\mathrm{fs}^{-1}$).}
\label{tab:toy_parameters}
\begin{tabular}{lccc}
Parameter & Dimer in WCA & Müller-Brown & 1D Potential \\
\hline
Dimension $d$ & $90$ & $2$ & $1$ \\
Time step $\Delta t$ & $0.005$ & $1.0\,\mathrm{fs}$ & $0.01\,\mathrm{fs}$ \\
Friction $\zeta$ & $1.0$ & $0.1\,\mathrm{fs}^{-1}$ & $2\,\mathrm{fs}^{-1}$ \\
AMS Replicas $N_T$ & $500$ & $1000$ & $200$ \\
AMS Realizations $n_{\mathrm{real}}$ & $50$ & $40$ & $40$ \\
\end{tabular}
\end{table}

\begin{table}[H]
\caption{Simulation parameters for the butane conformational analysis at $\mathcal{T} = 300\,\mathrm{K}$.}
\label{tab:butane_parameters}
\begin{tabular}{lc}
Parameter & Value \\
\hline
Time step $\Delta t$ & $1\,\mathrm{fs}$ \\
Friction coefficient $\zeta$ & $0.01\,\mathrm{fs}^{-1}$ \\
Reactant state $ A$ & $\phi^{-1}([175,185])$ \\
Product state $B$ & $\phi^{-1}([55,65] \cup [295,305])$ \\
AMS Replicas $N_T$ &  $100$ \\
AMS Realizations $n_{\mathrm{real}}$ & $80$ \\
\end{tabular}
\end{table}
\section{Simulations of butane conformations at different temperatures}\label{app:temperature}

\begin{figure}[H]
    \includegraphics[width=\linewidth]{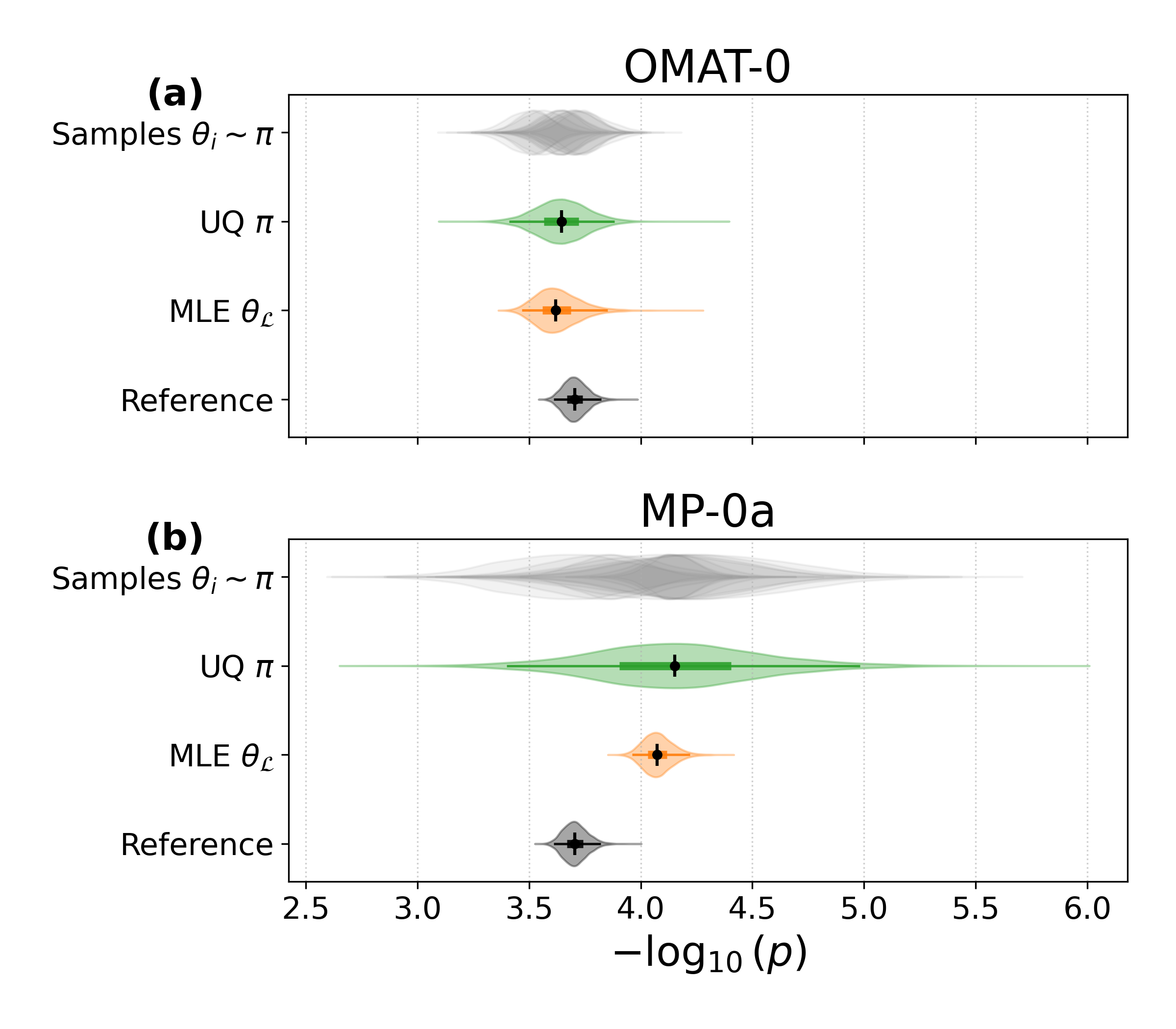}
    \caption{Violin plots of the probability density function of the negative log rare event probability of butane conformation, for \texttt{MACE-OMAT-0} (top) and \texttt{MACE-MP-0a} (bottom), at $\mathcal{T} = 200$~K. The reference distribution is shown in black, the prediction from the best-fit potential is in orange, and the full Bayesian uncertainty-aware distribution $p \mid \pi$ is in green. Samples of the parameter vector $\theta$ were drawn, and their corresponding conditional probabilities distributions are plotted in grey. }
    \label{fig:violin_200}
\end{figure}
\begin{figure}[H]
    \includegraphics[width=\linewidth]{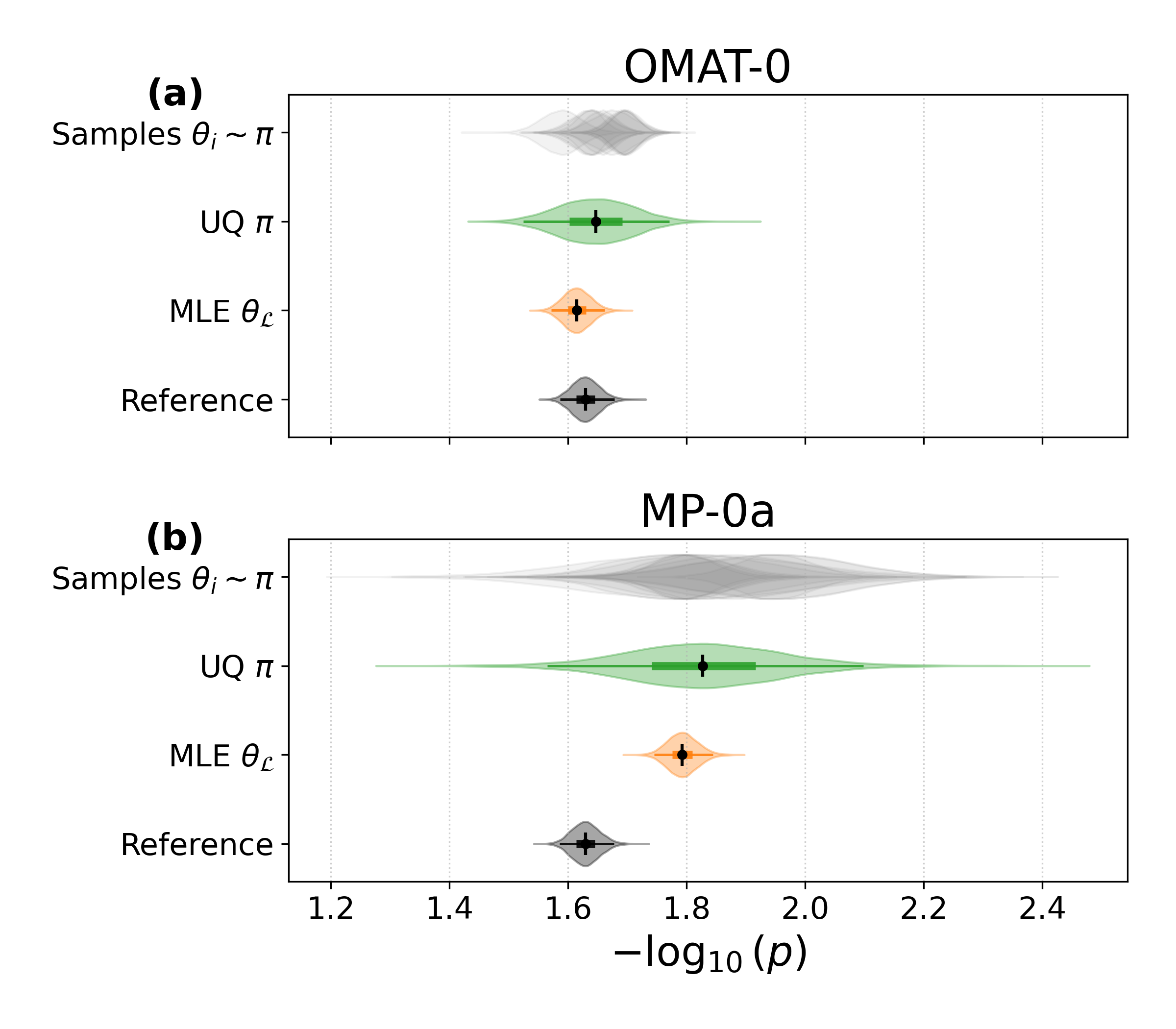}
    \caption{Violin plots of the probability density function of the negative log rare event probability of butane conformation, for \texttt{MACE-OMAT-0} (top) and \texttt{MACE-MP-0a} (bottom), at $\mathcal{T} = 500$~K. The reference distribution is shown in black, the prediction from the best-fit potential is in orange, and the full Bayesian uncertainty-aware distribution $p \mid \pi$ is in green. Samples of the parameter vector $\theta$ were drawn, and their corresponding conditional probabilities distributions are plotted in grey. }
    \label{fig:violin_500}
\end{figure}
\newpage
\nocite{*}

\bibliographystyle{iopart-num} 
\bibliography{aipsamp}

\end{document}